\documentclass[usenatbib]{mn2e}
\usepackage{natbib}
\usepackage{hyperref}
\usepackage{psfig}
\usepackage{epsfig}

\newif\ifAMStwofonts

\def\sqiglt{\hbox{\rlap{\lower.55ex \hbox {$\sim$}}\kern-.05em \raise.4ex \hbox{$<$}\,}}
\def\sqiggt{\hbox{\rlap{\lower.55ex \hbox {$\sim$}}\kern-.05em \raise.4ex \hbox{$>$}\,}}
\def\til{\ensuremath{\sim\,}}

\def\W{\ensuremath{\Omega}}

\def\chisq{\ensuremath{\chi^2}}
\def\rchisq{\ensuremath{\chi_{\nu}^{2}}}

\newcommand{\tim}[1]{\ensuremath{\times 10^{#1}}}
\def\deg{\ensuremath{^{\circ}}}
\def\etal{et al.\ }

\def\msol{\ensuremath{M_\odot}}

\def\cms{\ensuremath{$cm$^{-2}}}

\def\swift{\emph{Swift}}
\def\nh{\ensuremath{N_{\rm H}}}

\def\fermi{\emph{Fermi}}
\def\kw{\emph{Konus-Wind}}
\def\t0{\ensuremath{T_{0}}}
\def\W{\ensuremath{\mathcal{W}}}

\title[GRB 130925A]{GRB 130925A: an ultra-long Gamma Ray Burst with a dust-echo
afterglow, and implications for the origin of the ultra-long GRBs.}

\author[Evans et al.]{P.A. Evans$^1$\thanks{pae9@leicester.ac.uk}, R. Willingale$^1$, J.P. Osborne$^1$,
P.T. O'Brien$^1$, N.R. Tanvir$^1$ \and
D.D. Frederiks$^2$, V.D. Pal'shin$^2$, D.S. Svinkin$^2$, A. Lien$^{3,4,5}$, J. Cummings$^{3,4,5}$, \and 
S. Xiong$^6$, B.-B. Zhang$^6$, D. G\"otz$^7$, V. Savchenko$^8$, H. Negoro$^{9}$, \and 
S. Nakahira$^{10}$, K. Suzuki$^{9}$, K. Wiersema$^1$, R.L.C. Starling$^1$, \and 
A.J. Castro-Tirado$^{11,12}$, A.P. Beardmore$^1$,  R. S\'anchez-Ram\'{\i}rez$^{11}$, J. Gorosabel$^{11,13,14}$, \and
S. Jeong$^{11}$, J.A. Kennea$^{15}$, D.N. Burrows$^{15}$ and N. Gehrels$^3$
\\
\\
$^1$Department of Physics and Astronomy, University of Leicester, Leicester, LE1 7RH, UK \\
$^2$Ioffe Physical-Technical Institute, Politekhnicheskaya 26, St. Petersburg 194021, Russia \\
$^3$NASA Goddard Space Flight Center, Greenbelt, MD 20771, USA \\
$^4$Center for Research and Exploration in Space Science and Technology (CRESST), USA \\
$^5$Department of Physics and Center for Space Sciences and Technology, University of Maryland Baltimore County, Baltimore, MD 21250, USA \\
$^6$Center for Space Plasma and Aeronomic Research (CSPAR), University of Alabama in Huntsville, Huntsville, AL 35899, USA \\
$^7$Laboratoire AIM-Paris-Saclay, CEA/DSM/Irfu – CNRS – Universit\'e Paris Diderot, CE-Saclay,pt courrier 131, 91191 Gif-sur-Yvette, France \\
$^8$Fran\c{c}ois Arago Centre, APC, Universit\'e Paris Diderot, CNRS/IN2P3, CEA/Irfu, Observatoire de Paris, Sorbonne Paris Cit\'e, \\
\ \ 10 rue Alice Domon et L\'eonie Duquet,  F-75205 Paris Cedex 13, France \\
$^{9}$Department of Physics, Nihon University, 1-8-14 Kanda-Surugadai, Chiyoda-ku, Tokyo 101-8308  \\
$^{10}$ISS Science Project Office, Institute of Space and Astronautical Science (ISAS), Japan Aerospace Exploration Agency (JAXA), \\
\ \ 2-1-1 Sengen, Tsukuba, Ibaraki 305-8505 \\
$^{11}$Instituto de Astrof\'{\i}sica de Andaluc\'{\i}a (IAA-CSIC), Glorieta de la Astronom\'{\i}a s/n, E-18008, Granada, Spain \\
$^{12}$Unidad Asociada Departamento de Ingeniería de Sistemas y Autom\'atica, E.T.S. de Ingenieros Industriales, Universidad de M\'alaga, Spain \\
$^{13}$Unidad Asociada Grupo Ciencias Planetarias UPV/EHU-IAA/CSIC, Departamento de F\'{\i}sica Aplicada I, E.T.S., Ingenier\'{\i}a, \\
\ \ Universidad del Pa\'{\i}s Vasco UPV/EHU, Bilbao, Spain \\
$^{14}$Ikerbasque, Basque Foundation for Science, Bilbao, Spain \\
$^{15}$Department of Astronomy \&\ Astrophysics, The Pennsylvania State University, 525 Davey Lab, University Park, PA 16802, USA \\
}

\date{Accepted 
      Received }

\pagerange{\pageref{firstpage}--\pageref{lastpage}}
\pubyear{2013}

\begin{document}

\maketitle

\label{firstpage}

\begin{abstract} 
GRB~130925A was an unusual GRB, consisting of 3 distinct episodes of high-energy emission
spanning \til20 ks, making it a member of the proposed category of `ultra-long' bursts.
It was also unusual in that its late-time X-ray emission observed by \swift\ was very soft, and showed 
a strong hard-to-soft spectral evolution with time. This evolution, rarely seen in GRB
afterglows, can be well modelled as the dust-scattered echo of the prompt emission, with
stringent limits on the contribution from the normal afterglow (i.e.\ external shock) emission.
We consider and reject the possibility that GRB~130925A was some form of tidal disruption event,
and instead show that if the circumburst density around GRB~130925A is low, the long duration
of the burst and faint external shock emission are naturally explained. Indeed, we suggest that
the ultra-long GRBs as a class can be explained as those with low circumburst densities, such that the deceleration time
(at which point the material ejected from the nascent black hole is decelerated by the circumburst medium)
is \til20 ks, as opposed to a few hundred seconds for the normal long GRBs. The increased deceleration radius means that 
more of the ejected shells can interact before reaching the external shock, naturally explaining both the increased
duration of GRB 130925A, the duration of its prompt pulses, and the fainter-than-normal afterglow.
\end{abstract}


\section{Introduction}
\label{sec:intro}

Gamma-ray bursts (GRBs), discovered by \cite{Klebesadel73}, 
are the most powerful explosions in the universe. \cite{Mazets81} and
\cite{Kouveliotou93} showed that GRBs can be divided into two classes based on 
their duration: long and short GRBs. These objects have different progenitors,
with the short (\sqiglt2 s) GRBs believed to be the the mergers of binary neutron-star systems
and long GRBs arising from the collapse of a massive star (see \citealt{Zhang09a} for a detailed discussion
of GRB progenitors and classification). In both cases, it is generally believed
that the prompt emission arises due to interactions within the outflow of material (see, e.g.\ \citealt{Zhang07Review}).
Recently \cite{Gendre13}, \cite{Stratta13} and \cite{Levan14} have proposed an additional category of
`ultra-long' bursts, GRBs with durations of kiloseconds. These authors consider tidal disruption of
a white-dwarf star by a massive black hole, and a GRB with a  blue supergiant progenitor (larger than those
of normal long GRBs) as possible causes of these ultra-long bursts, with the latter being favoured.
In contrast, \cite{Virgili13} suggest that the ultra-long GRBs simply represent
the tail of the distribution of long GRBs.

With the exception of GRB~101225A, the ultra-long GRBs show an X-ray afterglow, once the 
prompt emission is over. Such a feature is seen after most long GRBs, and is generally believed
to occur when the material ejected by the GRB, which is travelling close to the
speed of light, is decelerated by the circumburst medium (CBM). A shock forms and propagates into the medium,
radiating by the synchrotron mechanism as it does so. This model is not uniformly accepted,
with some authors (e.g.\ \citealt{Uhm07,Genet07,Leventis13}) arguing that the late-time emission is strongly affected by emission from a reverse shock,
which propagates back into the out-flowing material once it is decelerated.

Regardless of their physical origin, GRB X-ray afterglows show a range of 
different light curve behaviours \citep{Evans09}, perhaps the most 
curious of which is the so-called `plateau' phase 
\citep{Nousek06,Zhang06}  -- a  period during which the afterglow fades slowly, if at all. The most 
widely-accepted explanation for this plateau is that there is an ongoing 
injection of energy into the shocked CBM \citep[e.g][]{Liang07}. Such plateaux are not seen in 
all afterglows: \cite{Evans09} found them in $<70\%$ of bursts.
In contrast to the light curves, the spectra of X-ray afterglows show little variation,
with the photon index ($\Gamma$; $N(E) dE\propto E^{-\Gamma}$) distribution\footnote{The XRT catalogue
quotes the spectral energy index, $\beta=\Gamma-1$} being approximately Gaussian,
with a mean of 2.0 and a FWHM of 0.7 (\citealt{Evans09}, the live XRT
GRB catalogue\footnote{http://www.swift.ac.uk/xrt\_live\_cat}). This spectrum is
generally found not to evolve with time \citep[e.g.][]{Butler07a,Shen09}.

In this paper we consider GRB~130925A, a GRB which triggered \swift, \fermi, \kw, \emph{INTEGRAL}, and
\emph{MAXI}, 
and had a duration of $>5$ ks, making it a candidate ultra-long GRB. However, this burst is also unusual
in that its late-time X-ray data showed a strong hard-to-soft spectral evolution with time. Recently,
\cite{Bellm14} have analysed \swift, \emph{Chandra} and \emph{NuSTAR} data of this burst, and claim
the presence of multiple afterglow components; however, we shall show that
a simpler emission model can explain the data presented here.

Throughout this paper we assume a cosmology with $H_0=71$ km s$^{-1}$ Mpc$^{-1}$, $\Omega_m=0.27, \Omega_{\rm vac}=0.73$,
and we made use of the online Cosmology Calculator\footnote{http://www.astro.ucla.edu/$\sim$wright/CosmoCalc.html} \citep{Wright06}.
Errors are at the 90\%\ level unless otherwise stated.

\begin{figure*}
\begin{center}
\psfig{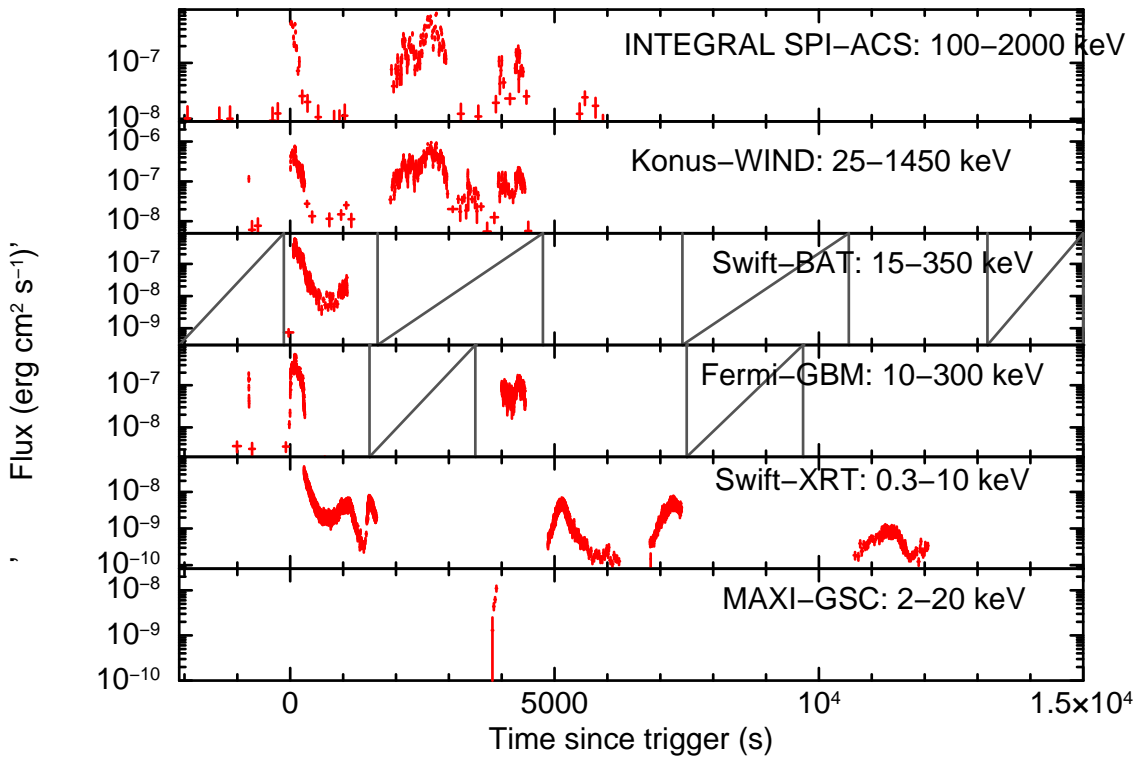}
\caption{Multi-observatory light curves of the prompt and flaring emission. These were built assuming a constant spectral model,
as fitted to the \emph{Episode 1} data (Section~\ref{sec:prompt}). The fluxes are given in each instrument's native band,
and in the observer frame. This reveals the relative flux at different energies, for each pulse, illustrating
the spectral variation from pulse to pulse.
The data have been binned to a minimum signal-to-noise ratio per bin of 5, using the
approach of \protect\cite{Evans10}. As \swift\ and \fermi\ are in low-Earth orbits,
the times when the source was outside of their field of view are marked by the grey diagonal lines.
For \swift-XRT whenever the source was in the field of view it was detected, so to keep the plot
simple we do not mark the times when it was not in the field (although these will be similar to the BAT times). Similarly for {\emph MAXI\/}
which could only observe the GRB for \til2 min of each \til93 min orbit (and only detected
the GRB in one orbit) we do not include the observability intervals.
}
\label{fig:earlyCurve}
\end{center}
\end{figure*}

\section{Observations}
\label{sec:obs}

GRB~130925A triggered the \emph{INTEGRAL} SPI-ACS instrument at 04:09:25 UT on 2013 
September 25 \citep{Savchenko13}; hereafter this time is referred to as \t0.
\fermi-GBM triggered just after this at  04:09:26.73 UT \citep{Fitzpatrick13,Jenke13}, 
and \emph{Swift}-BAT triggered at 04:11:24 UT \citep{Lien13}; the GRB was also detected by  \kw\ 
in waiting mode \citep{Golenetskii13}. These triggers all correspond
to the same episode of emission, which lasted around 900 seconds (in the 15--350 keV BAT data
the total duration above the background level was 846 s, while $T_{90}=179$ s). There was an earlier
`precursor' lasting ~6 s which triggered the \fermi-GBM at 03:56:23.29 UT ($\t0-781$ s) this was
also seen by \kw\ but not by \emph{INTEGRAL} or BAT. The \fermi\ trigger also resulted
in an automated slew of the satellite to orient the LAT boresight towards the GRB \citep{Jenke13};
however, no emission was detected in the 0.1--10 GeV band,
with an upper limit (95\%\ confidence) of 4.8\tim{-10} erg \cms\ s$^{-1}$
\citep{Kocevski13}.

The \swift-XRT began observing 147.4 s after the BAT trigger and found a bright, 
uncatalogued X-ray source \citep{Lien13}.

\begin{figure}
\begin{center}
\psfig{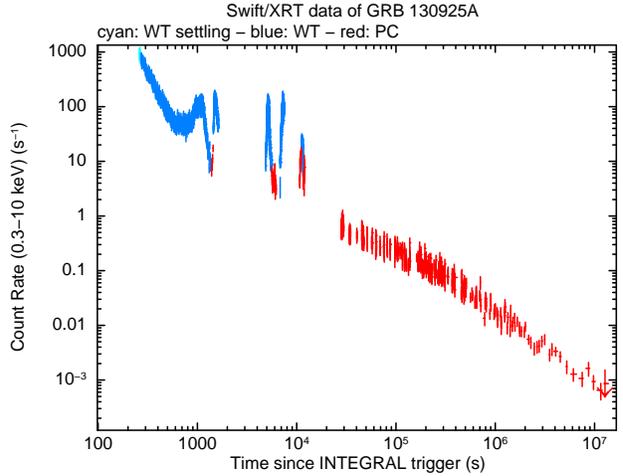}
\caption{The full 0.3--10 keV X-ray light curve, from the XRT light curve repository (Evans et al.\ 2009).}
\label{fig:xrtlc}
\end{center}
\end{figure}

A second episode of high-energy emission occurred at ~ $\t0+2000$--$3000$ 
s and was seen by both \kw\  and \emph{INTEGRAL}; the GRB was not 
observable by \swift\ or \fermi\ at this time due to Earth occultation. At 05:13:41 ($\t0+3.8$ ks) 
the \emph{MAXI} Gas Slit Camera also triggered  on the GRB 
\citep{Suzuki13} which still had a flux of 290 mCrab: this 
corresponds to the time of a third interval of high energy emission 
detected by  \kw, \emph{INTEGRAL} and \fermi-GBM (the object was 
outside the \swift-BAT field of view). As with the initial episode,
\fermi-LAT did not detect anything, with an upper limit of 1.6\tim{-9} erg \cms\ s$^{-1}$
(0.1--10 GeV, \citealt{Kocevski13}). At $\t0+4.8$ ks \swift\ 
observations resumed, and the XRT detected a flare which was also 
seen by the BAT, \emph{INTEGRAL} and GBM although at much lower levels
than from the three main emission episodes. Two further flares were detected by XRT on the subsequent 
spacecraft orbits\footnote{\swift\ has a \til96 min orbit.}, before the X-ray light curve settled down 
to the decay ubiquitous to X-ray GRB afterglows. 
Fig.~\ref{fig:earlyCurve} shows the multi-observatory light curve of the 
prompt emission and flaring episodes. For each instrument we obtained a single counts-to-flux
conversion factor using the joint spectral fit to the first emission episode (Section~\ref{sec:prompt})
and multiplied the count-rate by this value. This neglects the effects of
spectral evolution (which are, however, incoporated in the modelling in Section~\ref{sec:prompt})
but shows the relative strength of the various pulses in different energy bands.
The full XRT light curve (taken from the XRT light curve
repository\footnote{http://www.swift.ac.uk/xrt\_curves} [\citealt{Evans07,Evans09}]
on 2014 March 17) is given in Fig.~\ref{fig:xrtlc}. 

At longer wavelengths, an infra-red counterpart was detected by GROND
\citep{Greiner08} in  observations starting at \t0+567 s
\citep{Sudilovsky13}, and by RATIR \citep{Butler12} in observations
starting at \t0+8.28 ks \citep{Butler13}.  VLT spectroscopy found the
GRB redshift to be 0.347 \citep{Vreeswijk13} in agreement with our own
observations (Section~\ref{sec:optobs}). The \swift-UVOT did not detect
the burst; however, the IR colours from \cite{Sudilovsky13} suggest that
there is significant dust in the line of sight, consistent with the lack
of UVOT detection. Radio observations at 230 GHz beginning 1.1 d after
the trigger found no source, with a 3-$\sigma$ upper limit of 1.89 mJy
\citep{Zauderer13}, and observations at 93 GHz beginning 1.2 d after the
trigger also found no source, with a 3-$\sigma$ upper limit of 0.6 mJy.
Later radio observations taken with ATCA between \til15 and 21 days
after the trigger detected emission at the GRB location, with fluxes of
\til 140--190 $\mu$Jy at frequencies between 5.5 and 19 GHz
\citep{Bannister13}

Observations with the \emph{Hubble} Space Telescope (HST) revealed the host galaxy to
be a nearly edge-on spiral, but with  signs of disturbance, with the bulge being elongated perpendicular to the disk, suggesting that
the host is a polar ring galaxy. The afterglow was located in the HST images
to be 0.12$^{\prime\prime}$ offset from the centre of the galaxy, which is \til600pc in projection
\citep{Tanvir13}. HST observed the object again at two further epochs (Tanvir \etal in preparation).

\subsection{GTC imaging and spectroscopic observations of the GRB~130925A host galaxy}
\label{sec:optobs}

Imaging of the host galaxy of GRB~130925A in the $griz$ bands was carried out
with the 10.4 m GTC telescope equipped with the OSIRIS instrument on
the nights of 2013 Nov 4--5. The images were acquired in $2\times2$ binning, providing a 
pixel scale of 0.25$^{\prime\prime}$/pix. Photometric calibration
was performed by observation of standard star SA114$-$656
\citep{Smith02}. The images were dark-subtracted and flat-fielded using
custom {\sc iraf}\footnote{{\sc iraf} is distributed by the National 
Optical Astronomy Observatory, which is operated by the Association
of Universities for Research in Astronomy (AURA) under cooperative 
agreement with the National Science Foundation.} routines. Aperture
photometry was done using {\sc daophot} tasks implemented in 
{\sc iraf}. Table~\ref{tab:hostmags} displays the host galaxy AB 
magnitudes. The $g$-band magnitude was used to scale the flux of 
the host galaxy GTC spectrum (see Table \ref{tab:hostspec}).

A simple single stellar population fit to the integrated host magnitudes 
using \cite{Bruzual93} models, a \cite{Calzetti00} extinction curve
and redshift of $z=0.348$, gives acceptable fits 
for a young stellar population (\til30 Myr) and substantial extinction 
($A_V\sim2.2$ mag). However, we caution that the morphology 
of the host \citep{Tanvir13}, in particular the presence of a red 
bulge and blue disk (Tanvir \etal\ in prep.), indicates that more complex 
models may be required to characterize the host properties.

\begin{table}
\begin{center}
\begin{tabular}{cccc}
\hline\hline Observing Date & Exposure & Filter  & Magnitude \\
    (Start--End) 2013 UT & time (s)              &         &  (AB) \\
\hline
Nov 5.111541--5.116603  & $3\times120$             &  $g$    &
$22.72\pm0.08$ \\
Nov 4.083744--5.130114  & $4\times90+3\times60$    &  $r$    &
$21.94\pm0.05$ \\
Nov 5.117194--5.120173  & $3\times60$              &  $i$    &
$21.68\pm0.07$ \\
Nov 5.120764--5.126661  & $5\times75$              &  $z$    &
$21.16\pm0.07$ \\
\hline
\end{tabular}
\caption{Observing log of the host galaxy imaging. The magnitudes are
in the AB system with no reddening correction. The $r$-\
band measurement is based on data taken in two consecutive nights. Errors
are at the 1-$\sigma$ level.}
\label{tab:hostmags}
\end{center}
\end{table}
 
In addition spectral observations were carried out with the
GTC(+OSIRIS) on 2013 Nov 5, between 01:26 UT and 02:22 UT, with a total exposure time of
3$\times$900 s. The spectra were acquired with grism R1000B, providing a
spectral range of 3615--7760 \AA. The data were taken with a slit
width of $1.49^{\prime\prime}$, resulting in a resolution of $R\til550$
(estimated using weak sky lines). Data reduction followed standard 
procedures using custom routines under {\sc iraf} and python. The spectra were
bias-corrected and flat-fielded. We have chosen a wavelength solution
based on calibration arcs taken with a slit width of
$1.2^{\prime\prime}$ to achieve better accuracy than the one we also
obtained with the $1.49^{\prime\prime}$ one. The flux of the final
spectra were calibrated with the spectrophotometric standard 
G191-B2B \citep{Oke90} and scaled to the host galaxy $g$-band
magnitude (see Table~\ref{tab:hostmags}) to account for the slit losses.
We identified several lines in the spectrum, at a common redshift of \til0.348
(see Table~\ref{tab:hostspec}) which we adopt as the redshift of the GRB hereafter; this
gives a luminosity distance of 1.836 Gpc. 
We derive a lower limit on the star formation rate (SFR) from the strength
of the [O II] line, applying the calibration of \cite{Kennicutt98}, SFR
($M_{\odot}$/yr) = ($1.4 \pm 0.4$) \tim{-41}  $L_{\rm [O II]}$.
Using the measured line luminosity as a lower limit implies
SFR($M_{\odot}$/yr) $>$ 0.95 $M_{\odot}$/yr , a lower value than 
inferred from other GRB host galaxies \citep{Christensen04}).

\begin{table*}
\begin{center}
\begin{tabular}{cccccc}
\hline
Ion & $\lambda_{\rm obs}$ & $\lambda_{\rm rest}$ & $z$ & FWHM  & Observed flux  \\
    & (\AA\ air)    & (\AA\ air)  & & (\AA)  & (erg cm$^{-2}$ s$^{-1}$)  \\
\hline
[OII] & 5028.5 $\pm$ 0.1 & 3728.815 & 0.34855 & 9.7 $\pm$ 0.5 & $(1.68\pm0.09) \times 10^{-16}$ \\
$H_{\beta}$ & 6555.8 $\pm$ 0.3 & 4861.363 & 0.34855 & 9.7 $\pm$ 0.5 & $(7.1\pm0.4) \times 10^{-17}$ \\ 
${\rm [OIII]}$ & 6685.8 $\pm$ 0.5 & 4958.911 & 0.34824 & 10.2 $\pm$ 0.4 & $(3.9\pm0.1) \times 10^{-17}$ \\
${\rm [OIII]}$ & 6750.5 $\pm$ 0.2 & 5006.843 & 0.34825 & 10.2 $\pm$ 0.4 & $(1.16\pm0.04) \times 10^{-16}$ \\
\hline
\end{tabular}
\caption{Emission lines identified in the host galaxy of GRB~130925A, revealing the redshift to be \til0.348.
Errors are at the 1-$\sigma$ level.}
\label{tab:hostspec}
\end{center}
\end{table*}

\begin{figure}
\begin{center}
\psfig{file=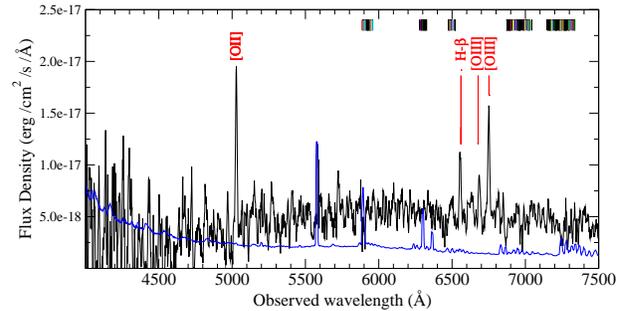,height=8.1cm,angle=-90}
\caption{The optical spectrum of GRB 130925A from the GTC. The blue line shows the level of the errors. The tick marks at the top
indicate the atmospheric sky lines/bands. Various
emission lines can be seen in the spectrum at a redshift of 0.348.}
\label{fig:optspec}
\end{center}
\end{figure}

\section{Prompt emission and flares}
\label{sec:prompt}

Due to the unusual duration of GRB~130925A, we examined whether the intervals of high energy emission
look like typical GRB prompt emission pulses (apart from their duration). Based on the light curve in Fig.~\ref{fig:earlyCurve}, we defined four intervals of high energy emission,
and extracted spectra for each of these from whichever instruments were on target at the time, as shown in Table~\ref{tab:hespectimes}\footnote{No
\emph{INTEGRAL\/} spectra were available due to the distance of the GRB from the satellite boresight}. For \fermi-GBM data a
spectrum was created individually for each detector which detected the source during the time interval. 

\begin{table*}
\begin{center}
\begin{tabular}{ccl}
\hline
Name & Times$^a$ & Instruments \\
\hline
Precursor & $-800$ to $-778$ & \fermi-GBM$^b$, \kw \\
Episode 1   & $-5$ to 300 & \fermi-GBM$^c$, \kw, \swift-BAT \\
Episode 2   & 1800 to 3000 & \kw \\
Episode 3   & 3800 to 4500 & \fermi-GBM$^d$, \kw \\
Flare 1     & 780 to 1200 & \swift-XRT and BAT \\
Flare 2     & 1200-1400   & \swift-XRT and BAT \\
Flare 3     & 4750-5350   & \swift-XRT and BAT \\
Flare 4     & 6680-7270   & \swift-XRT and BAT \\
Flare 5     & 10530-11590   & \swift-XRT \\
\hline
\end{tabular}
\caption{Times of the prompt emission episodes, over which high-energy spectra were 
extracted, and the 5 X-ray flares for which \swift\ spectra were obtained. We also note which missions and instruments
gathered spectroscopic data during each episode.\newline
$^a$ Times in seconds since \t0. $^b$ Data from 4 NaI detectors. $^c$ Data from 1 BGO detector
and 2 NaI. $^d$ Data from 3 NaI detectors.}
\label{tab:hespectimes}
\end{center}
\end{table*}

We fitted the spectra of these time intervals in {\sc xspec} \citep{Arnaud96} with three models: a 
power-law, cut-off power-law, and Band function \citep{Band93}. For each fit the 
parameters were tied to be the same for all instruments, but a  
multiplicative normalisation factor was allowed to vary between them to allow 
for  calibration differences in the absolute flux level. For the 
precursor the cut-off power-law and Band models offered 
no significant improvement over the simple power-law. For the other 
spectra the cut-off power-law was significantly better than the simple power-law. The Band function offered no further improvement,
tending towards unconstrained highly negative values for the high-energy index, at which point the Band function behaves
as a cut-off power-law. 
The best fitting spectral parameters for the cut-off power-law and Band model fits are given in Table~\ref{tab:hespecfits}.

\begin{table*}
\begin{center}
\begin{tabular}{ccccccccc}
\hline
 & &  \multicolumn{3}{c}{Cut-off power-law} & \multicolumn{4}{c}{Band function} \\
Name & Fluence  & Photon index  & $E_{\rm peak}$  & \chisq\ ($\nu$) & $\Gamma_{\rm low}$ & $\Gamma_{\rm high}$ & $E_{\rm peak}$ (keV) & \chisq\ ($\nu$)  \\
 & (erg cm$^{-2}$) & ($\Gamma$) & (keV)  & & & & (keV) \\
 & (15--350 keV) \\
\hline
Precursor & 6.8\tim{-7} & $2.06^{+0.28}_{-0.21}$  & --- & 604 (511)$^a$ \\
Episode 1 & 8.0\tim{-5} & $1.91\pm0.03$ & $65^{+13}_{-16}$ & 670 (436) & $1.91\pm0.03$ & $>2.9$ & $65^{+13}_{-16}$ & 670 (435)  \\
Episode 2 & 3.8\tim{-4} & $1.55^{+0.04}_{-0.05}$ & $175^{+13}_{-10}$ & 10$^{-5}$ (0)$^b$ \\
Episode 3 & 6.0\tim{-5} & $1.58^{+0.12}_{-0.13}$ & $94^{+14}_{-10}$ & 410 (363) & $1.57^{+0.12}_{-0.13}$ & $>2.9$ & $94^{+14}_{-11}$ & 410 (362) \\
\hline
\end{tabular}
\caption{Details of the spectral fits to the episodes of prompt emission. \newline
$^a$The precursor pulse was best fitted as a simple power-law. \newline
$^b$ The \kw\ spectrum, which is the only one available for this episode, contains only 3 bins. Even so, the cut-off power-law
is very clearly a much better fit to the data (for the power-law fit, \chisq=380.5 for $\nu=1$);
however, it also has 0 degrees of freedom so a \rchisq\ value cannot be produced. We did not fit the Band
model to this spectrum as it has -1 degrees of freedom.}
\label{tab:hespecfits}
\end{center}
\end{table*}

We also created spectra covering the five flares that are seen in the XRT light 
curve (Table~\ref{tab:hespectimes}). For the first four spectra we have both Windowed Timing (WT) mode 
XRT data and BAT data (taken in survey mode). Although the source was 
not detected by BAT during the second flare the data provide 
constraints. The final flare was too faint for BAT to make a meaningful 
contribution, but we have both WT and Photon Counting (PC) mode data for 
that flare.  Following the latest calibration 
guidance\footnote{http://www.swift.ac.uk/analysis/xrt/digest\_cal.php},
as this source is moderately absorbed we used only single pixel (grade 
0) events and ignored the data below 0.6 keV. We used the gain files and RMF from the 2013-04-20
release of the \swift-XRT CALDB\footnote{http://heasarc.gsfc.nasa.gov/docs/heasarc/caldb/swift}.
A turn-up was seen in the WT data 
below 0.8 keV, which could not be modelled even by adding thermal 
components to the spectra, and we therefore treated these as residual 
calibration systematics (which will be modelled in forthcoming calibration 
releases) and excluded them from the fits. The XRT spectra were fitted 
using the {\sc xspec} w-statistic
\footnote{http://heasarc.nasa.gov/xanadu/xspec/manual/\newline
XSappendixStatistics.html} (\W; i.e.\ requesting the C-stat but 
supplying a background spectrum), while the BAT spectra were fitted at 
the same time using the \chisq\ statistic. The fit results are 
shown in Table~\ref{tab:xrtspecfits}; the absorption used was a {\sc 
phabs} component fixed to the Galactic value of $1.7\tim{20}$ \cms\ 
\citep{Willingale13} with a {\sc zphabs} component with the redshift fixed 
at 0.348, and the column density free to vary overall, but tied to the same value for all flares. Note that, as with the prompt pulses,
flare spectra tend to evolve through the flare, thus our fits give average values.


\begin{table*}
\begin{center}
\begin{tabular}{lcccccccc}
\hline
Name & Time$^a$ & \multicolumn{3}{c} {Power-law} &  \multicolumn{4}{c} {Cut-off power-law} \\ 
     &      & \nh ($10^{22}$ \cms) &  $\Gamma$ & F-stat$^b$ (dof)  &  \nh ($10^{22}$ \cms) &  $\Gamma$ & $E_{\rm cut}$ (keV) & F-stat$^b$ (dof)  \\
\hline
Flare 1 & 901--1321    & $1.86\pm0.03$ & $1.65\pm0.03$ & 4397 (4148)  & $1.75\pm0.03$ & $1.57\pm0.03$ & $68^{+66}_{-23}$ & 4317 (4143) \\
Flare 2 & 1321--1626   & --"-- & $1.76\pm0.04$  & --"-- & --"-- & $1.00\pm0.16$ & $3.90^{+0.32}_{-0.24}$ & --"-- \\
Flare 3 & 4872--5472   & --"--  & $2.06\pm0.03$  & --"-- & --"--  & $1.92^{+0.05}_{-0.06}$ & $3.7^{+2.1}_{-1.3}$ & --"-- \\
Flare 4 & 6672--7391    & --"--  & $1.66\pm0.02$ & --"-- & --"-- & $1.55^{+0.03}_{-0.04}$ &  $23^{+13}_{-7}$  & --"-- \\
Flare 5 & 10650--11710 & --"--  & $2.35\pm0.05$ & --"-- & --"-- & $1.93^{+0.09}_{-0.06}$ &  $0.509^{+0.018}_{-0.017}$   & --"-- \\
\hline
\end{tabular}
\caption{Details of the spectral fits to the 5 flares seen in the X-ray light curve. The flares were
fitted simultaneously, with the absorption free to vary overall, but tied to be the same for all flares.
\newline $^a$ Seconds since \t0. $^b$ i.e. the total fit-statistic, $F=\chisq+\mathcal{W}$.}
\label{tab:xrtspecfits}
\end{center}
\end{table*}

\subsection{Pulse modelling}
\label{sec:pulsemodel}

The spectral fits above give the average spectra of the emission episodes, but the spectrum varies 
between pulses and within each pulse (which is why \chisq\ is often large). Thus to properly
consider the prompt emission we need to model the data in a way that includes both spectral and brightness
variation with time. We did this using the pulse modelling technique of \cite{Willingale10}. This models the \swift-BAT light curve (in four energy bands)
and/or the XRT light curve (in two energy bands) of each individual pulse or flare with a functional model. The model
defines how the brightness and spectrum of the flare evolves with time, and depends on the 
peak time of the flare (since the trigger), $T_{pk}$, the time since the flaring material was
ejected by the central engine, $t_f$, and the spectrum of the flare. The latter
is a Band function whose peak energy decays as $t^{-1}$ after the flare peak time.
The later-time XRT data are also modelled, with the afterglow component described in Section~\ref{sec:extshock}.
To fit this model to the BAT data we use look-up tables created for the 
standard BAT energy bands; however, the BAT only collected event-mode 
data during the first sequence of pulses in the interval \t0+56 s to \t0+319 
s. We therefore used the \kw\ data, which covers the entirety of the prompt emission.
We mapped \kw\ band 1 (25--95 keV) to 
BAT bands 1 (15--25 keV) and 2 (25--50 keV) and \kw\ bands 2+3 (95--1450 
keV) to BAT bands 3 (50--100 keV) and 4 (100--350 keV) to provide 
reasonable energy overlap and good statistics. We normalised the 
combined \kw\ rates to match the individual BAT band rates over the 
overlap time interval \t0+56 s to \t0+319 s, within which the pulse 
structure observed by BAT and \kw\  are identical. The resulting
BAT-energy-band light curves contain a combination of BAT and \kw\ data. For 
the later pulses these combined light curves are exclusively \kw\ data, 
renormalised using the scaling factors from the first sequence of 
pulses. The scaling factors will be correct providing the average 
spectrum doesn't change significantly. Spectral fitting results 
are shown in Table~\ref{tab:hespecfits}. The photon index varies from 1.5 to 1.9 
and the peak energy from 65 to 175 keV. These 
differences introduce changes of 10-20\% in the scaling factors over the 
4 BAT energy bands, which are small compared with the typical uncertainties on 
the individual data points. The spectrum used in the pulse fitting of the light curves had
a fixed cut-off energy of 370 keV\footnote{The cut-off energy, $E_c$ is related to the
peak energy $E_p$ by $E_p=E_c(2-\Gamma)$. Formally, the fit was a Band function,
with the high energy index set to $-10$, as in \cite{Willingale10}; however, 
this model is effectively the same as a cut-off power-law, and so is consistent with the spectral fits.}
(equivalent to 500 keV in the source frame) and gave a mean pulse photon index of 1.9.

\begin{figure*}
\begin{center}
\vbox{
  \psfig{file=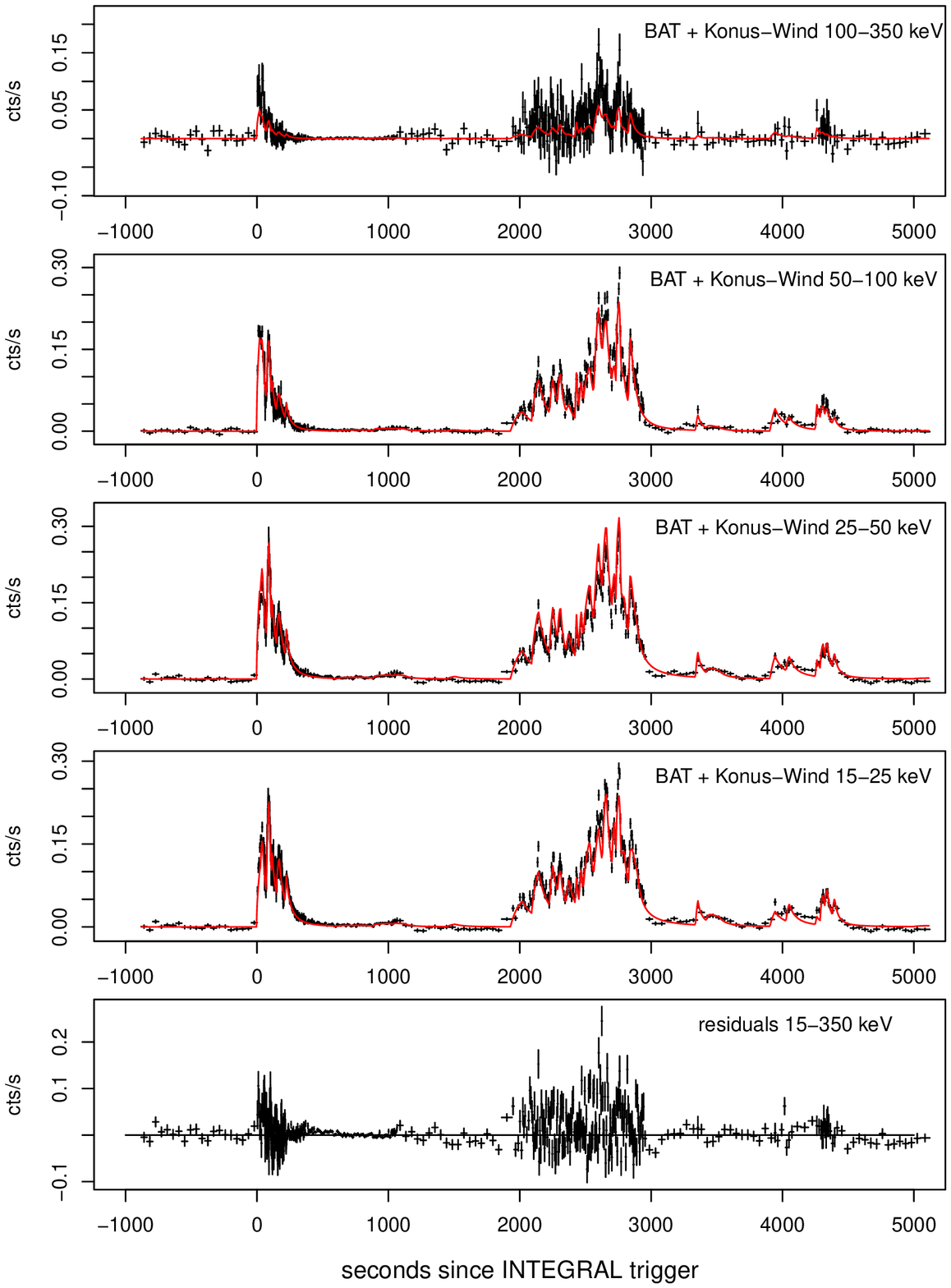,width=12cm}
  \hspace{-0.4cm}\psfig{file=fig4b.eps,height=12.2cm,angle=-90}
  \vspace{0.5cm}
}
\caption{\emph{Top 4 panels:} The BAT+\kw\ data for the prompt emission in the standard BAT bands, along
with the fitted pulse model (red) from \protect\cite{Willingale10} and residuals. While some fine detail of the pulses
are not perfectly fitted, the basic shape, time and spectral behaviour of the pulses
are well reproduced by our model. The count-rates are normalised to the equivalent BAT values in count s$^{-1}$ per detector
values.
\emph{Bottom panel:} The \kw\ hardness ratio of counts in the hardest
to softest band. Data were binned to a minimum signal-to-noise ratio of 5 in each band,
and the data points with large errors during the quiescent periods were removed.
The spectral evolution can be clearly seen.}
\label{fig:promptmod}
\end{center}
\end{figure*}

The data and fitted models are shown in Fig.~\ref{fig:promptmod}, with the fit parameters given
in Table~\ref{tab:promptfit}. While the model does not match all of the pulses in detail
(\rchisq=3.3 for 3869 degrees of freedom) the basic shape, time and spectral shape of the pulses are well reproduced.
The peak bolometric (1--10$^4$ keV) isotropic luminosity of the prompt emission derived from this modelling is
$L_{\rm iso}=4.5\pm0.6$\tim{50} erg s$^{-1}$, occurring at \t0+22 s; integrating
over the pulses we find the total bolometric isotropic fluence $E_{\rm iso}=2.9\pm0.3$\tim{53} erg.

\begin{figure}
\begin{center}
\psfig{file=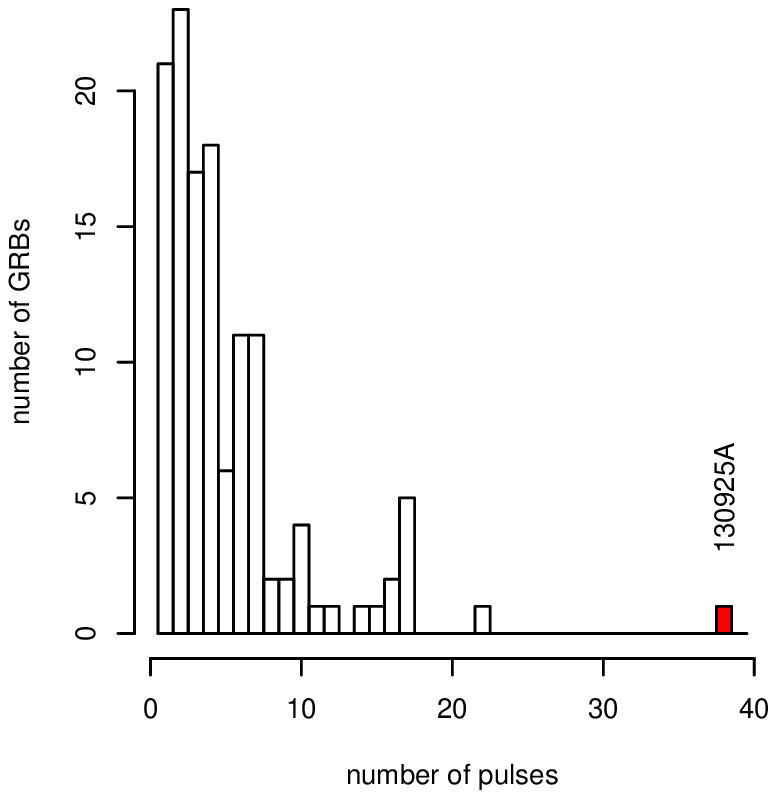,height=5.1cm,angle=-90}
\psfig{file=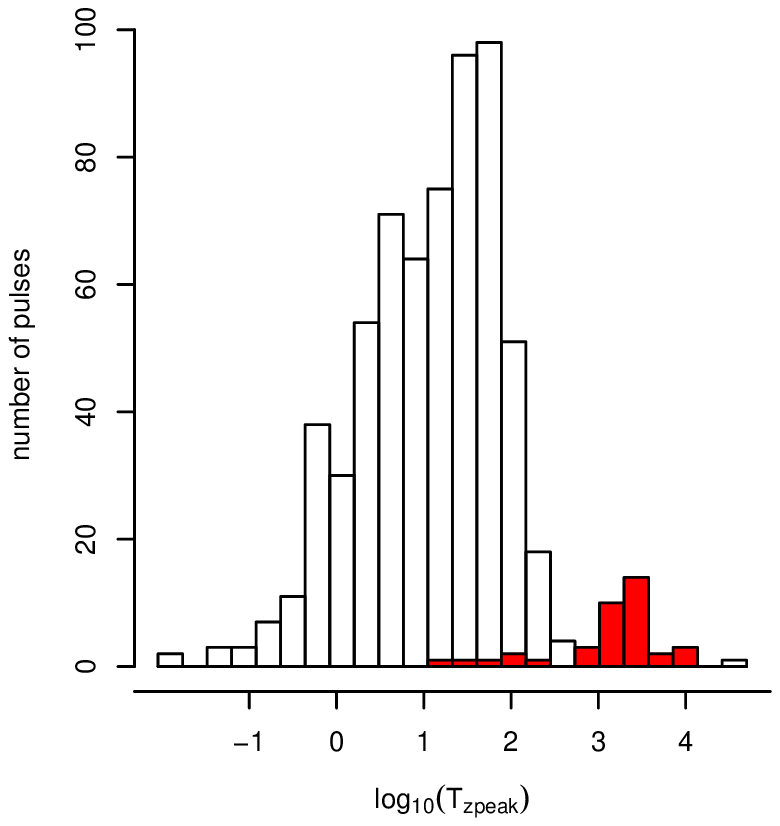,height=5.1cm,angle=-90}
\psfig{file=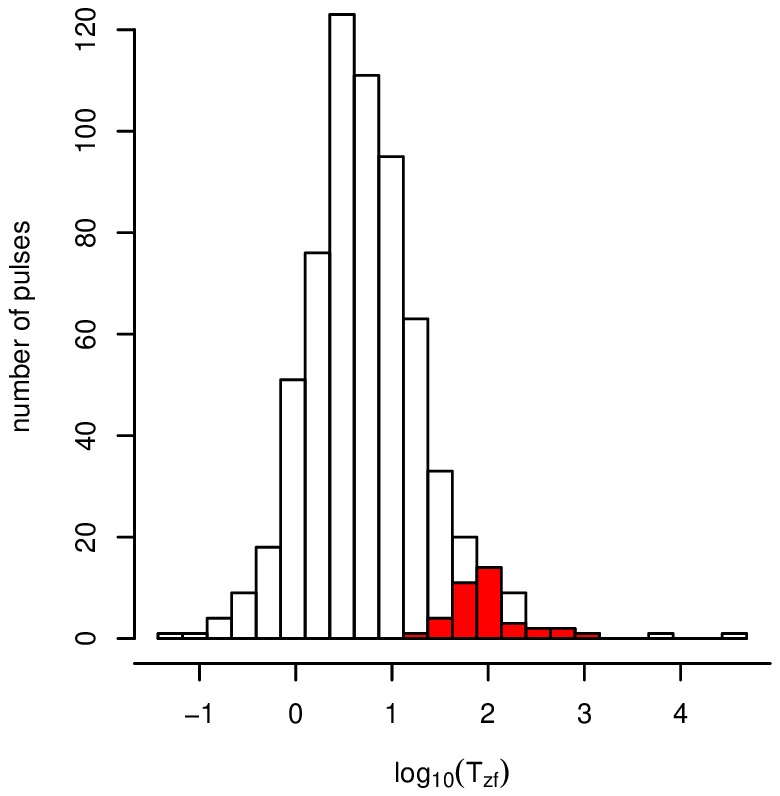,height=5.1cm,angle=-90}
\caption{Comparison of the prompt emission properties of GRB~130925A with the 127 GRBs with known redshift observed by 
\swift-BAT and XRT up to 2011 May. GRB~130925A is in red.
\emph{Top}: The distribution of the number of pulses needed to model the prompt emission.
\emph{Middle}: The distribution of the peak time of the pulses in the GRBs' rest frame.
\emph{Bottom}: The distribution of the duration of the pulses in the GRBs' rest frame.
The number of pulses and their peak times are unusually large compared to the population of
GRBs as a whole. The pulse durations in GRB 130925A are at the high end of the overall
distribution, although not inconsistent with the general range.}
\label{fig:pulse1}
\end{center}
\end{figure}

\begin{figure}
\begin{center}
\vbox{
\psfig{file=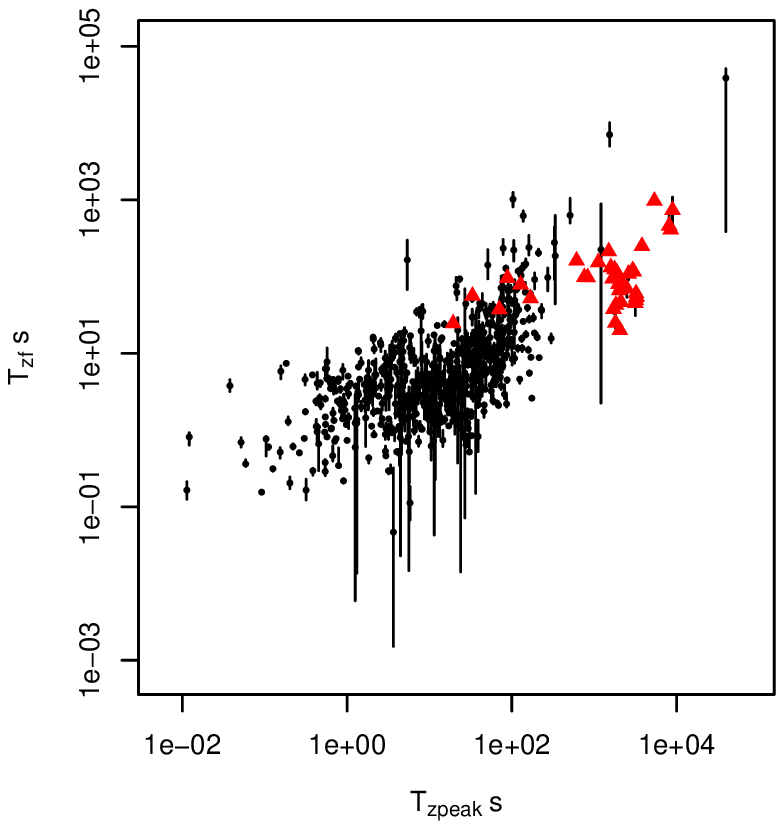,height=8.1cm,angle=-90}
\vspace{0.4cm}
\psfig{file=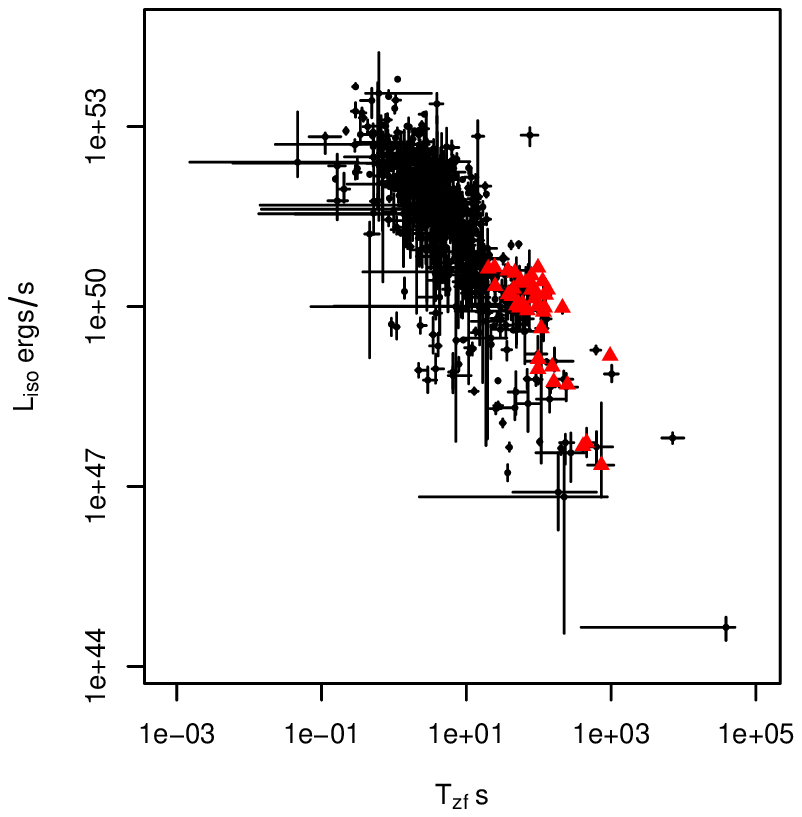,height=8.1cm,angle=-90}
}
\caption{Comparison of the prompt emission relationships of GRB~130925A with the 127 GRBs with known redshift observed by 
\swift-BAT and XRT up to 2011 May. GRB 130925A is in red.
\emph{Top}: The pulse duration plotted against the pulse peak time (both in the GRBs' rest frames); GRB~130925A lies along
the correlation seen for the population at large.
\emph{Bottom}: The isotropic-equivalent luminosity of the pulses against the pulse duration (rest frame). The pulses for
GRB~130925A tend to be longer for their luminosity (i.e. more energetic) than the generality of GRB~pulses.}
\label{fig:pulse2}
\end{center}
\end{figure}

Since the publication of \cite{Willingale10}, one of us (RW) has fitted the BAT pulses and XRT flares for 127 GRBs with a redshift and early XRT data
up to 2011 May, so we compared the results for GRB~130925A with that sample (which does not include any of the other ultra-long GRBs). GRB~130925A required
38 distinct pulses, substantially more than any other GRB in our sample (Fig.~\ref{fig:pulse1}, top). Not surprisingly given the duration 
of GRB~130925A, most of these pulses peak at a rest-frame time much later than the generality of GRB pulses (Fig.~\ref{fig:pulse1}, middle);
also the pulses are longer (in the GRB rest frame) than most prompt pulses, although within the distribution found
from the population at large (Fig.~\ref{fig:pulse1}, bottom). For the pulse population as a whole, a correlation is seen between
the rest-frame $T_{\rm pk}$ and $T_f$ values (the pulse peak time and duration respectively, Fig.~\ref{fig:pulse2}, top),
and an anti-correlation exists between the rest-frame duration
and the isotropic-equivalent peak luminosity of the pulses (Fig.~\ref{fig:pulse2}, bottom). 
As Fig.~\ref{fig:pulse2} shows, the pulses in GRB~130925A are consistent with 
the first of these correlations, but are a factor of \til5--10 more luminous for their durations than is typical for GRB pulses.
In summary, the prompt emission pulses are largely consistent with what we see
in most GRBs, except that there are more of them, extending to later times than normal, and they carry more energy than typical pulses
of the same duration.

\begin{table*}
\begin{center}
\begin{tabular}{rccccccc}
\hline
Pulse \# & $T_{pk}$ (s) & $T_{f}$ (s) & 90\% conf range & $\Gamma^1$ & 90\% conf range  &  $L_{iso}$ (erg s$^{-1}$) & 90\% conf range \\
\hline
1 & 22     & 33 & 31 -- 37 & 1.19 & 1.12 -- 1.27 & 4.52\tim{50} & 3.98\tim{50} -- 5.10\tim{50} \\
2 & 41     & 76 & 72 -- 80 & 1.71 & 1.63 -- 1.80 & 2.68\tim{50} & 2.38\tim{50} -- 3.05\tim{50} \\
3 & 91     & 50 & 49 -- 53 & 1.90 & 1.86 -- 1.93 & 3.97\tim{50} & 3.63\tim{50} -- 4.84\tim{50} \\
4 & 115    & 129 & 118 -- 139 & 1.95 & 1.77 -- 2.07 & 1.02\tim{50} & 7.96\tim{49} -- 1.39\tim{50} \\
5 & 168    & 106 & 103 -- 115 & 2.11 & 2.06 -- 2.14 & 1.90\tim{50} & 1.27\tim{50} -- 2.12\tim{50} \\
6 & 223    & 70 & 67 -- 75 & 2.07 & 2.02 -- 2.12 & 1.14\tim{50} & 9.71\tim{49} -- 1.30\tim{50} \\
7 & 820    & 217 & 210 -- 225 & 1.63 & 1.58 -- 1.68 & 5.53\tim{48} & 4.69\tim{48} -- 6.47\tim{48} \\
8 & 1020   & 133 & 132 -- 134 & 1.73 & 1.70 -- 1.74 & 1.34\tim{49} & 1.24\tim{49} -- 1.44\tim{49} \\
9 & 1120   & 131 & 127 -- 135 & 1.86 & 1.81 -- 1.91 & 9.08\tim{48} & 8.08\tim{48} -- 1.03\tim{49} \\
10 & 1508  & 207 & 202 -- 215 & 2.11 & 2.05 -- 2.16 & 9.93\tim{48} & 9.19\tim{48} -- 1.08\tim{49} \\
11 & 2020  & 287 & 267 -- 315 & 1.81 & 1.56 -- 1.92 & 9.50\tim{49} & 7.68\tim{49} -- 1.23\tim{50} \\
12 & 2143  & 178 & 167 -- 191 & 1.63 & 1.54 -- 1.71 & 1.92\tim{50} & 1.74\tim{50} -- 2.13\tim{50} \\
13 & 2252  & 127 & 117 -- 139 & 1.73 & 1.59 -- 1.86 & 1.66\tim{50} & 1.40\tim{50} -- 2.00\tim{50} \\
14 & 2311  & 51 & 45 -- 58 & 1.43 & 1.24 -- 1.63 & 1.59\tim{50} & 1.28\tim{50} -- 2.00\tim{50} \\
15 & 2374  & 167 & 142 -- 198 & 2.36 & 2.26 -- 2.43 & 1.56\tim{50} & 5.27\tim{49} -- 2.84\tim{50} \\
16 & 2432  & 34 & 29 -- 44 & 1.18 & 0.95 -- 1.46 & 2.19\tim{50} & 1.48\tim{50} -- 3.02\tim{50} \\
17 & 2469  & 58 & 51 -- 67 & 1.69 & 1.52 -- 1.91 & 1.50\tim{50} & 1.18\tim{50} -- 2.04\tim{50} \\
18 & 2532  & 153 & 147 -- 158 & 1.80 & 1.65 -- 1.86 & 2.64\tim{50} & 2.32\tim{50} -- 3.02\tim{50} \\
19 & 2599  & 132 & 127 -- 139 & 1.26 & 1.20 -- 1.32 & 4.44\tim{50} & 4.11\tim{50} -- 4.81\tim{50} \\
20 & 2658  & 107 & 103 -- 114 & 1.92 & 1.81 -- 1.98 & 3.30\tim{50} & 2.78\tim{50} -- 3.85\tim{50} \\
21 & 2719  & 89 & 83 -- 98 & 2.24 & 2.08 -- 2.35 & 2.49\tim{50} & 1.63\tim{50} -- 4.49\tim{50} \\
22 & 2760  & 27 & 27 -- 28 & 1.49 & 1.43 -- 1.55 & 4.30\tim{50} & 3.97\tim{50} -- 4.65\tim{50} \\
23 & 2795  & 119 & 108 -- 131 & 1.95 & 1.80 -- 2.25 & 1.37\tim{50} & 9.45\tim{49} -- 2.26\tim{50} \\
24 & 2842  & 63 & 59 -- 74 & 1.19 & 1.09 -- 1.32 & 3.64\tim{50} & 2.88\tim{50} -- 4.19\tim{50} \\
25 & 2895  & 100 & 90 -- 113 & 2.78 & 2.43 -- 2.65 & 2.73\tim{50} & 4.23\tim{49} -- 8.49\tim{50} \\
26 & 3356  & 94 & 72 -- 132 & 1.96 & 1.48 -- 2.31 & 8.62\tim{49} & 4.88\tim{49} -- 1.79\tim{50} \\
27 & 3517  & 146 & 122 -- 181 & 2.34 & 1.77 -- 3.01 & 4.32\tim{49} & 2.45\tim{47} -- 1.19\tim{50} \\
28 & 3943  & 162 & 150 -- 183 & 1.28 & 1.07 -- 1.51 & 9.63\tim{49} & 7.42\tim{49} -- 1.29\tim{50} \\
29 & 4050  & 157 & 143 -- 172 & 2.25 & 1.91 -- 2.68 & 8.16\tim{49} & 2.68\tim{49} -- 2.16\tim{50} \\
30 & 4261  & 61 & 42 -- 85 & 0.69 & 0.21 -- 1.24 & 1.96\tim{50} & 9.11\tim{49} -- 4.41\tim{50} \\
31 & 4309  & 67 & 60 -- 80 & 2.02 & 1.59 -- 2.28 & 1.05\tim{50} & 6.87\tim{49} -- 2.17\tim{50} \\
32 & 4339  & 81 & 68 -- 92 & 2.22 & 1.77 -- 2.59 & 1.11\tim{50} & 3.59\tim{49} -- 2.93\tim{50} \\
33 & 4396  & 73 & 62 -- 86 & 2.31 & 1.84 -- 2.99 & 9.57\tim{49} & 4.34\tim{49} -- 4.69\tim{50} \\
34 & 5120  & 336 & 325 -- 343 & 2.37 & 2.33 -- 2.41 & 4.98\tim{48} & 4.72\tim{48} -- 5.32\tim{48} \\
35 & 7259  & 1305 & 1287 -- 1323 & 1.73 & 1.67 -- 1.77 & 1.50\tim{49} & 1.37\tim{49} -- 1.73\tim{49} \\
36 & 10970 & 619 & 531 -- 830 & 2.62 & 2.29 -- 2.93 & 5.37\tim{47} & 3.12\tim{47} -- 9.36\tim{47} \\
37 & 11439 & 551 & 527 -- 574 & 2.78 & 2.66 -- 2.90 & 4.70\tim{47} & 3.89\tim{47} -- 5.72\tim{47} \\
38 & 12036 & 989 & 637 -- 1464 & 2.57 & 0.28 -- 3.50 & 2.27\tim{47} & 6.65\tim{46} -- 2.47\tim{48} \\
\hline
\end{tabular}
\caption{The best-fitting parameters for the 38 pulses. $T_{peak}$ was not fitted but set by eye. Times are in 
the observer frame.\newline$^1 \Gamma$ is the spectral photon index of the pulse, this is constant for that pulse,
whereas $E_{\rm peak}$ evolves with time. See Willingale \etal(2010) for details.}
\label{tab:promptfit}
\end{center}
\end{table*}


\section{The spectrally evolving X-ray afterglow}
\label{sec:ag}

\begin{figure}
\begin{center}
\psfig{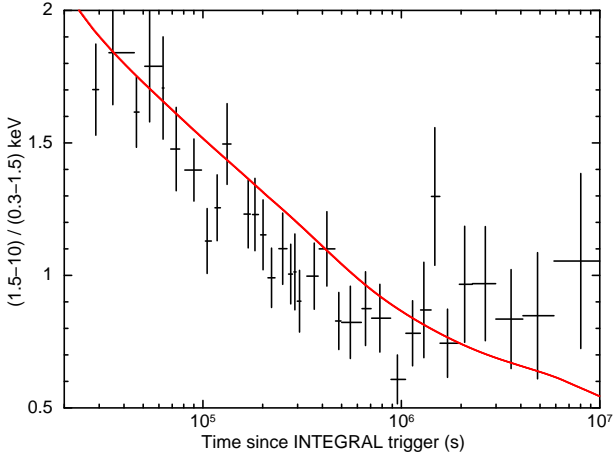}
\caption{\swift-XRT hardness ratio time series, showing the ratio of counts in the 1.5--10 keV and 0.3--1.5 keV bands.
The data shown begin at \t0+20 ks (i.e.\ once the prompt emission and flaring had ceased). The strong hard-to-soft evolution can be clearly seen.
The red line shows the hardness ratio predicted by the dust-scattering model (Section~\ref{sec:dustscat}).}
\label{fig:hr}
\end{center}
\end{figure}

GRBs show a wide variety of X-ray afterglow behaviour; however, one thing they all
have in common is that almost no evidence for late-time spectral evolution has been reported\footnote{the exception being GRB 090417B, which will be discussed later} (e.g.\ \citealt{Butler07a,Evans09}).
However, the XRT hardness ratio of GRB~130925A, after the flaring behaviour has subsided, shows a strong spectral evolution
from \t0+20 ks to \t0+\til700 ks (Fig.~\ref{fig:hr}). Fitting the hardness ratio time series from \t0+20 ks with a broken power-law
(i.e. HR $\propto t^{-\zeta}$ up to the break, after which the HR is constant) yielded a fit
with \chisq=23.2 ($\nu=31$). The break time, where the evolution ceased, is 
(8.3$^{+2.1}_{-2.6}$)\tim{5} s, and $\zeta=0.256^{+0.030}_{-0.026}$ (errors at 1-$\sigma$)
i.e. the source is getting softer with 10-$\sigma$ significance!
A similar behaviour has been reported in one previous burst: GRB~090417B for which the late-time X-ray data was interpreted
by \cite{Holland10} as scattering of the prompt emission off a dust screen, rather than emission from an external shock. 

We attempted to model the late-time\footnote{i.e. $t>20$ ks, after all of the X-ray flaring and prompt emission has finished}
X-ray emission of GRB~130925A in two ways: first as an external shock, and then using dust scattering.

\subsection{The X-ray afterglow as an external shock}
\label{sec:extshock}

To model the afterglow as the external shock, we followed
\cite{Willingale10}, combining the results of the pulse modelling with 
the functional form of the afterglow flux evolution developed by
\cite{Willingale07}, which consists of an exponential relaxing to a
power-law. The latter is fitted simultaneously to the 0.3--1.5 and
1.5--10 keV XRT light curves. When a late-time break was added to the model
($t_{\rm break}=3.4^{+2.5}_{-0.7}\tim{2}$ ks), this was able to reproduce the shape
of X-ray light curve from \t0+\til20 ks, but  some form of spectral evolution
had to be included in order to properly model the evolution
simultaneously in the 0.3--1.5 and 1.5--10 keV bands.  We therefore
modelled the spectrum as a power-law, whose photon index evolved with
time as

\begin{equation}
\Gamma = \Gamma_0 * \left(\frac{t}{t_a}\right)^\xi.
\end{equation}

\noindent until the late break, at which point the evolution ceased\footnote{The spectral evolution 
probably ends slightly later than the light curve break; however, we equate the two to
limit the number of free parameters.}. As noted in Section~\ref{sec:pulsemodel}, this was fitted simultaneously with the pulse model,
and yielded \rchisq=3.3 for 3869 degrees of freedom; most of the \chisq\ contribution comes from the prompt modelling.
The fit gave $\Gamma_0=-2.12^{+0.8}_{-0.5}$, $t_a=18.9^{+9.6}_{-6.4}$ ks (= the start of the afterglow plateau phase,
as in the \citealt{Willingale07} model) and $\xi=0.067^{+0.066}_{-0.094}$. This value encompasses 0 (i.e.\ no spectral evolution)
which implies that the spectral evolution is not significant; however, this is an artefact of
the number of free parameters and the correlations between them. For example, if we fix the time of
the late break, and the temporal decay after this break (features constrained by the 
light curve) the 90\%\ confidence interval for $\xi$\ becomes 0.037--0.089. Further,
if we perform the fit with all parameters free except for $\xi$, and fix $\xi$=0 (i.e.\ no spectral
evolution), \chisq\ increases by 18.4; an F-test therefore shows the evolution to be necessary at
the \til98\%\ level.

\begin{figure}
\begin{center}
\psfig{file=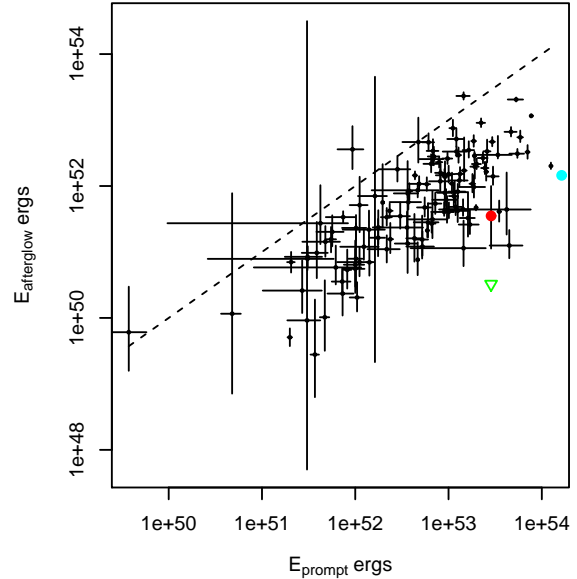,height=8.1cm,angle=-90}
\caption{The distribution of afterglow fluence against prompt fluence for the long GRBs in our sample. The red point shows the afterglow fluence
of GRB~130925A: the $E_{\rm afterglow}/E_{\rm prompt}$ is lower than for most bursts.
The green triangle is the upper limit on external shock emission in the dust-scattering model. In this case, 
the external shock emission must be significantly lower, as a fraction of the prompt emission, than for any other GRB.
The cyan point is the ultra-long GRB~121027A.
}
\label{fig:promptag}
\end{center}
\end{figure}

In the best-fit model (with spectral evolution), the isotropic-equivalent 0.3--350 keV peak (i.e. at $t=t_a$)
luminosity of the afterglow is $L_{\rm ag}=5.3^{+9.7}_{-3.6} \tim{46}$ erg s$^{-1}$
and the total 0.3--350 keV fluence of the afterglow is $3.5^{+6.5}_{-2.4} \tim{51}$ erg (this is measured by integrating
the model over all times). This means GRB~130925A has one of the lowest ratios
of afterglow to prompt fluence seen in the sample of 127 GRBs analysed (see Fig.~\ref{fig:promptag}).

In order to investigate in more detail possible physical causes of the spectral evolution, we extracted a series of spectra 
between $\t0+27.8$ ks and $\t0+2000$ ks (i.e.\ from the first XRT snapshot after the flaring had ended until the spectral evolution
had stopped), producing one spectrum every 250 accumulated counts, giving 27 spectra in total. We then fitted these spectra simultaneously
in {\sc xspec}. We initially fitted an absorbed power-law, with two photoelectric absorption components. The first
was a {\sc phabs} fixed at the Galactic value of 1.7\tim{20} \cms, the second was
a {\sc zphabs} with a redshift fixed at 0.348, and the column density free, but tied between the 27 spectra (i.e. time-invariant).
The power-law photon index and normalisation were free parameters. The best fit gave $\W=4122$, 
for 4703 degrees of freedom. This spectrum has no physical interpretation within the
synchrotron model, but serves as a baseline to compare other models with. These fits
showed no evidence for the high-energy residuals reported by \cite{Bellm14}.

We next tried replacing the power-law with a broken power-law, with the photon index above the break fixed to be 0.5 higher than
the photon index below the break. Only the low-energy slope, break energy and normalisation were allowed to vary between the fits. 
This reproduces the spectral evolution expected if the synchrotron cooling frequency is moving through the XRT bandpass. 
This gave a worse fit than the power-law fit ($\W=4426$, $\nu=4758$) and the break energy was extremely variable, showing no sign of the
steady evolution expected of the synchrotron cooling frequency,

We also tried fitting a power-law plus blackbody, to investigate whether some evolving optically thick component
could be present and modifying the fit \citep[e.g. ][]{Starling13,Campana06}. In this model the power-law photon index was tied between spectra; we used a {\sc zbbody}
model (i.e. a blackbody, with the temperature set in the GRB rest frame) with the redshift fixed at 0.348. The best fit gave $\W=4255$ ($\nu=4675$), again this is worse than simply having an 
evolving power-law. Furthermore, the blackbody temperature was highly variable with no steady evolution and frequently it tended to extreme values
(i.e. $10^{-4}$ or 200 keV: the model limits). 

Since this paper was posted on arXiv, \cite{Piro14} have also published an analysis of the data, in which they claim the detection of
blackbody emission during this interval of strong spectral evolution, in contrast to our result above. However, they fitted a single
\swift\ spectrum (`A1' in their paper) covering the interval \t0+20--300 ks, during which the spectrum evolves significantly
(Fig.~\ref{fig:hr}); whereas we used multiple spectra (with good S/N) during this interval. Fitting a single,
non-evolving component to a strongly evolving spectrum sometimes results in spurious 
extra components being needed to reproduce the spectrum, but these are artefacts of the inadequate model. Our approach of time-slicing during this
strong evolution is less prone to such effects, thus we reiterate our quantitative result from the previous paragraph:
the spectral evolution observed in this burst cannot be modelled as
a constant-spectral power-law with an evolving blackbody.

In summary: to model the late-time X-ray emission as arising from an external shock, we need to add a late-time
break, and we need to impose spectral evolution, the physics of which we cannot account for with the confines
of the external shock model: we therefore suggest than an alternative explanation is needed for the late-time X-ray data.

\subsection{The X-ray afterglow as dust scattering}
\label{sec:dustscat}

Scattering of X-rays from a GRB by dust in our galaxy has been detected previously \citep{Vaughan04}. 
The formation of an afterglow by the scattering of prompt X-rays by dust in the host galaxy was
considered by \cite{Klose98} and modelled by \cite{Shao07}, who
were able to reproduce the morphology of X-ray afterglow light curves. This work was then extended by \cite{Shen09} who considered
the spectral predictions of the dust model (see also \citealt{Shao07}) and found that dust scattering causes the afterglow to
get softer with time, in contrast with observations. One counter-example is GRB~090417B, which does show significant softening during the afterglow, 
and \cite{Holland10} modelled that GRB using the dust scattering model. Here, we follow the same methodology to consider whether the
spectral evolution of GRB~130925A (which is significantly stronger than that of GRB~090417B) could be the result of dust scattering.

To do this, we took the prompt pulse model from Section~\ref{sec:pulsemodel} and for each pulse estimated the fluence 
as a function of energy, $S(E)$. We then assumed that all of this fluence was emitted at a single
moment in time at $T_{pk}$ of that pulse, and calculated the flux which is scattered off a dust screen towards the observer. For a delay time
after each pulse, $t_{s}=t-T_{pk}$, the echo flux expected for a given photon energy, $E$, and dust grain size, $a$, is given by

\begin{equation}
F_{E,a}(t_{s})= \frac{S(E)}{t_{s}} \tau(E,a,t_{s}),
\end{equation}

\noindent where $\tau(E,a,t_{s})$ is the scattering optical depth. Because the
scattering occurs in the host galaxy at redshift $z$ we express the optical depth 
using parameters in the rest frame of the host. The 
scattering angle, $\theta$ is related to the distance of the dust from the
GRB, $R_{s}$, and the delay time in the observed frame, $t_s$:
$\theta=\sqrt{2ct_{s}/((1+z)R_{s})}$.
We can separate out the angular dependence of the optical depth
using the spherical Bessel function of the first order,
$j_{1}(x)=sin(x)/x^{2}-cos(x)/x$, giving:

\begin{equation}
\tau(E,a,t_{s})=2\tau_{a}(a,E)j_{1}^{2}(x(E,a,t_{s})).
\end{equation}

\noindent The rest-frame wavelength of observed photon energy $E$ is $\lambda=hc/(E(1+z))$ 
and $x=2\pi a\theta/\lambda$ is the scaled scattering angle.
Using the Rayleigh-Gans approximation dependence of $\tau_{a}(a,E)$ on the energy and grain size 
is given by

\begin{equation}
\tau_{a}(a,E)=\tau_{o}\left(\frac{E(1+z)}{1keV}\right)^{-2}
\left(\frac{a}{0.1\mu m}\right)^{4-q}
\end{equation}

\noindent where the grain size distribution is $dN(a)/da\propto a^{q}$. The
normalisation $\tau_{o}$ is the optical depth of the dust layer
at 1 keV for a grain size of 0.1 $\mu$m.
The total echo from a single layer of dust at distance $R_{s}$
at the observed energy $E$
is obtained by integrating over the grain size distribution

\begin{equation}
F_{E}(t_{s}) = \int_{a_-}^{a_+} F_{E,a}(t_{s}) {\rm d}a.
\end{equation}

The afterglow model of GRB~130925A was generated by summing the echoes
from every pulse in the prompt fit and folding the resultant
spectrum through the {\em Swift}-XRT response to produce predicted count
rate light curves in 2 energy bands, 0.3--1.5 keV and 1.5--10.0 keV.
We used a \chisq\ fit to find the best parameters. To allow a distribution of
dust along the line of sight, the dust was treated as being
in a sequence of 10 evenly spaced layers starting at a minimum distance
of $R_{m}$ pc and stretching over a radial range $R_{r}$ pc with a total optical
depth specified by $\tau_{o}$ as described above.
The grain size distribution index $q$ and dust grain
size limits $a_-$ and $a_+$ $\mu$m were included in the search.
The total optical depth of the dust column at energy $E$ is given by the
integral over dust grain size
$\tau_{s}(E)=\int_{a_-}^{a_+} \tau_{a}(E) {\rm d}a$ using the best fit value for
$\tau_{o}$.

The quality of the fit to the multi-band light curve using the dust scattering model for the afterglow
was about the same as that achieved using the standard afterglow model (Section~\ref{sec:extshock}): there were
120 free parameters (1 less than the standard model)
with 3990 data points giving $\chi^{2}_{\nu}=3.36$ (this includes
the contribution from the pulse model fit to the prompt data).
The best fit values and $90\%$ confidence
ranges for all the fitted dust parameters are given in Table~\ref{tab:dust}.
As $\tau_0$ is slightly greater than unity, the single-scattering approximation we have used
is not strictly valid; however, the impact of this simplification is expected to be small.

\begin{table}
\begin{center}
\begin{tabular}{lcc}
\hline
Parameter & Value & Error range \\
\hline
$\tau_{0}$        & 1.16 & 1.10-1.35 \\
$a_-$ $\mu$m      & 0.021 & 0.0001-0.040 \\
$a_+$ $\mu$m      & 0.285 & 0.250-0.400 \\
$q$               & 5.0   & 4.6-5.8 \\
$R_{m}$ pc        & 77    & 72-175 \\
$R_{r}$ pc        & 2000  & 1060-3250 \\
\hline
\end{tabular}
\caption{The best fitting parameters to model the late-time X-ray emission as dust scattering of the prompt emission.}
\label{tab:dust}
\end{center}

\end{table}

\begin{figure}
\begin{center}
\psfig{file=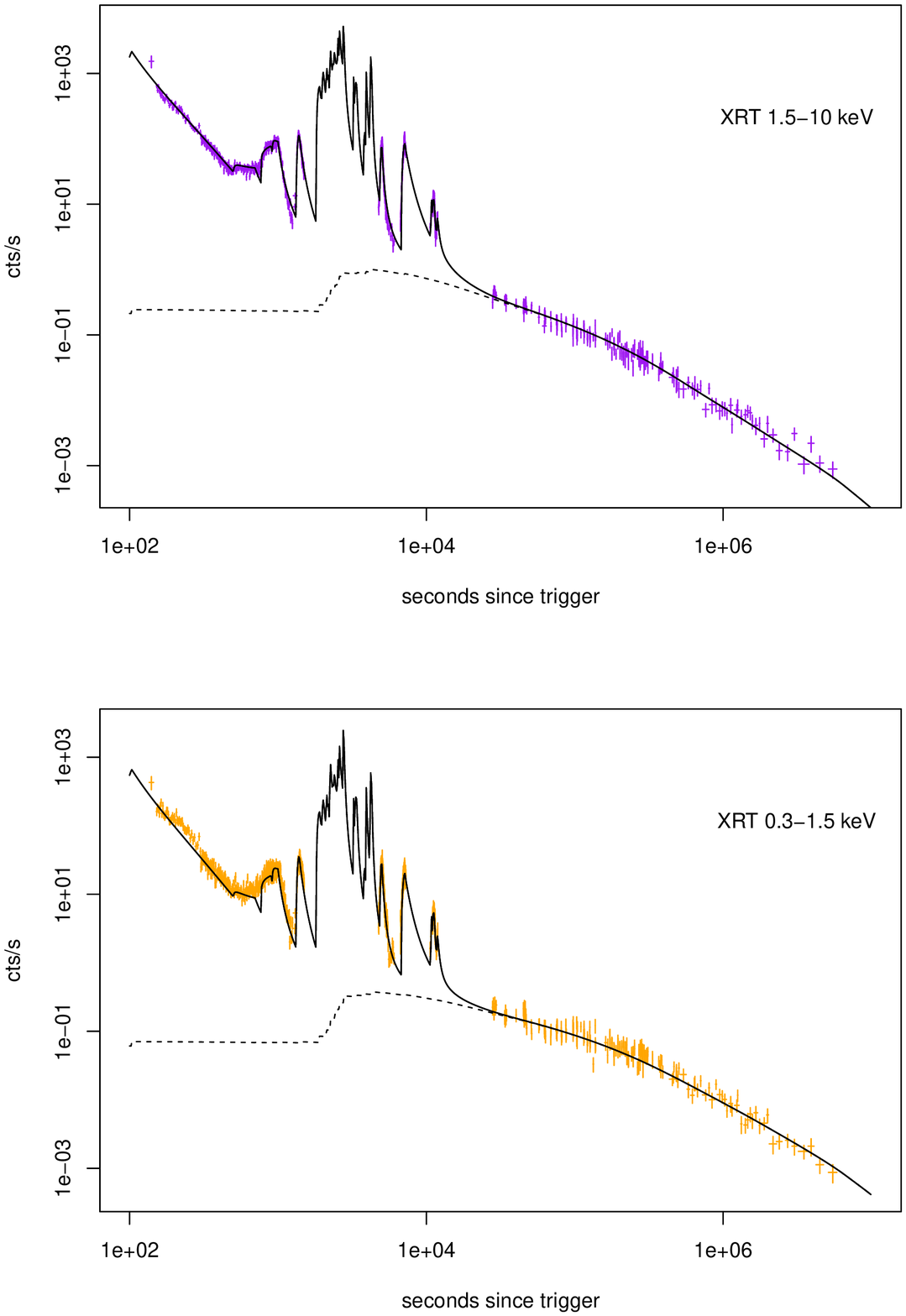,width=8.1cm,angle=0}
\end{center}
\caption{The dust model fit to the late-time XRT data GRB~130925A. The solid line shows the
model previously fitted to the prompt emission, plus the dust model. The dust model is shown as the dashed line.
The top and bottom panels show the hard and soft XRT bands respectively, illistrating
the good fit of the dust models to both bands.}
\label{fig:dustfit}
\end{figure}

\begin{figure}
\begin{center}
\psfig{file=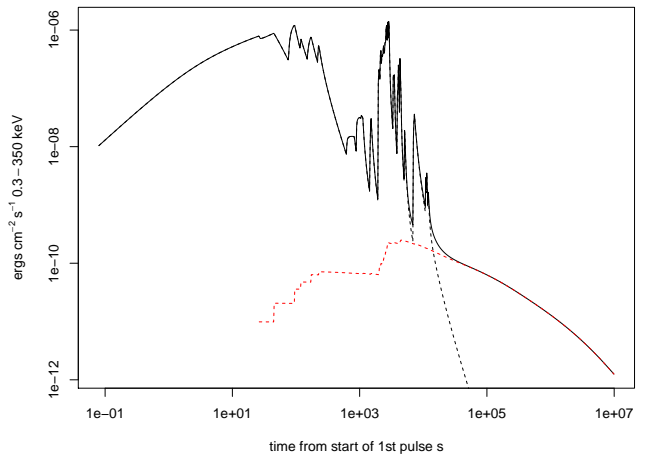,width=8.1cm,angle=0}
\end{center}
\caption{The best-fitting prompt emission and dust scattering model, in flux units over the 0.3--350 keV band.
The stepping behaviour in the rise of the dust echo shows the injection of each prompt pulse,
which is treated as instantaneous.}
\label{fig:dustmod}
\end{figure}

Whereas for the external shock model we had to articially add a late break and spectral
evolution to the model in order to fit the data, the dust scattering model fits all the pertinent features of the afterglow
naturally: the luminosity of the plateau, the initial slow decay from the plateau,
the soft spectrum at the start of the decay and the evolution of the
spectrum during the decay and the late break (Figs.~\ref{fig:dustfit} \&\ \ref{fig:dustmod}).

The combination of these
features provides a useful constraint on all the fitted parameters. The
optical depth, $\tau_{s}$ and the upper size limit, $a_+$ dominate the
plateau and early decay behaviour while the lower size limit, $a_-$ and
index $q$ set the overall decay. The 90\% range for $a_-$ indicates
an upper limit and, not unreasonably, that the grain size distribution probably
extends down to very small values.
The best fit value for the size
index, $q=5$, we derived here is significantly larger than the canonical
value of $q=3.5$ usually adopted \citep{Mathis77}, although
\cite{Predehl95} find a median value of $q=4.0$ from
analysis of dust scattering halo distributions observed in our Galaxy.
The upper limit to the grain size, $a_+=0.29$ $\mu$m is consistent
with values obtained in similar studies \citep[e.g.\ ][]{Predehl95,Holland10}.
The minimum distance, $R_{m}$ and
radial spread, $R_{d}$ set the curvature and position of the late break seen in 
the light curve at \til80 ks
The fitting clearly favours a distribution of dust along the line of sight,
with a depth of at least 1 kpc, rather than a single thin dust layer.
Furthermore, the model approximately reproduces fairly well the correct spectral index
and spectral evolution for the afterglow of GRB~130925A.

Figure \ref{fig:dustfit} shows the fitted XRT light curves. Figure \ref{fig:dustmod}
shows the model 0.3--350 keV flux for both the prompt and afterglow component
from the start of the burst through to the final decay.

The stepping behaviour of the rise of the dust echo arises because we have included
every prompt pulse individually. After each pulse an approximately
constant flux is added to the dust echo. The echo flux from each
pulse then starts to decay at a characteristic time after the pulse given by
\cite{Shen09}.

\begin{equation}
t_{c}=4.5\times10^{4}
\left(\frac{E}{1keV}\right)^{-2}
\left(\frac{R}{100pc}\right)
\left(\frac{a}{0.1\mu m}\right)^{-2}
{\rm s} .
\end{equation}

\noindent We note that the analysis of \cite{Holland10}, who modelled the 
afterglow of GRB~090417B using essentially the same dust scattering model
reproduced the correct spectral evolution in the afterglow but was unable to predict the
spectral index correctly. In their model all of the prompt emission was approximated by a single $\delta$
function with an average spectrum. The current results were obtained using a more detailed
model for the prompt emission (a $\delta$ function for each prompt pulse) and by fitting to two XRT energy bands simultaneously.

We can estimate the expected optical extinction, $A_{V}$, using the 
relation given by \cite{Draine04}, $\tau_{s}/A_{V}\approx 0.15 (E/1 
keV)^{-1.8}$, and we can further estimate the associated total hydrogen 
column using the relation derived by \cite{Willingale13} for our Galaxy, 
$N_{Htot}/A_{V} = 3.2\times10^{21}$ cm$^{-2}$. These give $A_{V}=7.7$ 
mag and $N_{Htot}=2.5\times10^{22}$ cm$^{-2}$. Both these relationships 
were derived using data from the Milky Way but there is substantial 
evidence that the dust properties of GRB hosts are different from the Milky Way
or galaxies in our neighbourhood (see the discussion in \citealt{Shen09});
an SMC-like metallicity would give $A_{V}\til6.2$.
Despite these caveats the value of $N_{Htot}$ derived from the dust echo 
afterglow model is comparable to the intrinsic 
$N_{H}=(1\pm0.1)\times10^{22}$ cm$^{-2}$ at $z=0.348$ derived from the 
late-time XRT spectrum. Thus the dust required to produce the observed 
afterglow by X-ray scattering alone is consistent with the intrinsic 
absorbing column required to fit the X-ray spectrum. Also note that
the galaxy-integrated colours are consistent with a dusty galaxy
(Section~\ref{sec:optobs}). 

If substantial dust is present near the GRB, we may expect to 
observe evidence of dust destruction. According \cite{Waxman00} dust 
destruction occurs out to radii of about 10 pc from the GRB, while 
\cite{Fruchter01} suggested that X-ray effects can destroy dust out to 
radii of \til100 pc. According to Table~\ref{tab:dust}, the dust screen 
in GRB130925A extends from \til80--2000 pc; thus we expect only a small 
amount, if any, of the dust to be destroyed, and that at the inner edge 
of the screen: any visible signature of this is likely to be weak and 
attenuated by its passage through the screen. Note that, should any dust 
destruction occur, this would reduce the optical extinction along the 
line of sight, but not the absorption column inferred from X-rays.

\subsection{The X-ray afterglow as an external shock and dust scattering}
\label{sec:agboth}

While the dust emission appears to fit the observed late-time data we expect there to be some
contribution from an external shock, unless the circumburst medium is of an abnormally low
density. We thus added a standard afterglow component (Section~\ref{sec:ag}) to the dust model.
The time of the plateau start (i.e.\ $t_a$) was fixed at 18.9 ks (as obtained in the fit without dust): values earlier than this
cannot be constrained due to the brightness of the prompt emission. The photon index
of the standard afterglow was fixed at 2.0, the median value obtained for all afterglows
fitted by \cite{Willingale10}. The best-fit was obtained with no external shock component.
The inclusion of any emission from this component increased \chisq, because the spectrum
of the external shock was much harder than that observed (which is well reproduced by the dust model).
The peak afterglow flux permitted by the fit at the 90\%\ confidence level was 7.04\tim{-12} erg \cms\ s$^{-1}$ (at \t0+18.9 ks).
Integrating this external shock component over all times gives us a 90\%\ confidence upper
limit of $E_{\rm iso,afterglow}<3.3\tim{50}$ erg
for the total fluence of the external shock\footnote{Although the afterglow start time is not
known, moving this to earlier times changes the fluence by only \til1--2\%, as this occurs
very early compared to the duration of the afterglow.}. This is plotted against the prompt fluence
as a green triangle in Fig.~\ref{fig:promptag}, which shows that the energy radiated in the external shock, as 
fraction of the prompt energy, is lower than seen for any other GRB.

We therefore consider it likely that the X-ray `afterglow' emission from GRB~130925A
is in fact the prompt emission being scattered into our line of sight by dust in the GRB host galaxy,
rather than emission from the standard external shock seen in typical GRBs.

\subsection{Spectral evolution in other GRBs}
\label{sec:evolveOther}

Strong spectral evolution has now been found in the afterglows of GRBs 090417B and 130925A.
To investigate how widespread this phenomenon is, we systematically studied all GRB afterglows detected
by \swift-XRT up to GRB~131002A for which the observations had a time base of at least 20 ks.

\begin{figure}
\begin{center}
\psfig{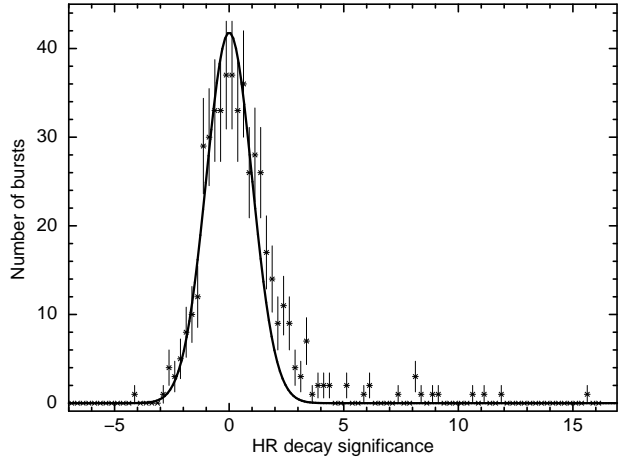}
\end{center}
\caption{The distribution of the significance, in $\sigma$, of any hardness ratio variation, for 672 XRT GRB afterglows
up to GRB~131002A. There is an excess of objects showing hard-to-soft spectral evolution; we investigated
those with $>5 \sigma$ significance in more detail.}
\label{fig:hrhist}
\end{figure}

We excluded the first 3 ks after the trigger (where the data may be affected by the 
prompt and high-latitude emission) and the times of any flares 
identified by the automatic fitting in the online XRT catalogue\footnote{http://www.swift.ac.uk/xrt\_live\_cat} \citep{Evans09};
we then fitted a power-law to the hardness ratio time series. For each fit we 
calculated the significance of the power-law index deviation from 0 (i.e. 
$\zeta/\sigma_\zeta$, where HR $\propto t^{-\zeta}$); a histogram of these values is given in 
Fig.~\ref{fig:hrhist}. There is an excess of objects 
with a spectral softening over time present at the \til2-$\sigma$ level, 
and 16 objects with evolution seen at the 5-$\sigma$ level.
We manually examined all of the latter; in five cases we found that the evolution was
caused either by flares which had not been adequately filtered out, or by a poorly sampled 
hardness ratio, where a single errant bin was dominating the fit.
However, bona fide spectral evolution was found in GRBs 130907A, 110709A, 100621A, 090404, 090417B, 090201, 081221, 080207 and 060218, as well
as GRB~130925A\footnote{There was also evidence for evolution in GRB~111209A, which is another ultra-long GRB.
However, in this case the light curve is apparently dominated by prompt, high latitude and flare emission until around 
$10^5$ s after the trigger. Fitting only the data after this time, the significance of the evolution reduces
to 1.5 $\sigma$.}. For these GRBs we created a series of spectra, starting a new one every \til250 counts, and 
fitted them with an absorbed power-law with the absorption component fixed, in a manner analogous 
to what we did for GRB~130925A in Section~\ref{sec:extshock}. For some of these GRBs the spectral evolution seen 
in the hardness ratio did not begin until part way through the light curve, and a broken power-law gave
a better fit to the HR evolution; in those cases we only took spectra from the time of the break onwards.

\begin{figure}
\begin{center}
\psfig{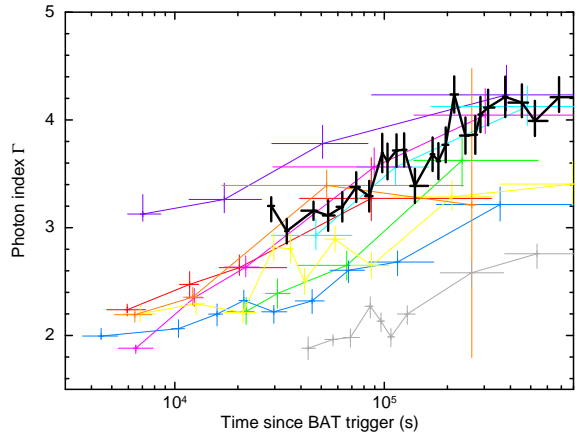}
\end{center}
\caption{The spectral photon index as a function of time, for the GRB afterglows which show spectral softening.
The photon index is derived from fitting absorbed power-law  models to a series of time-resolved spectra.}
\label{fig:gammatime}
\end{figure}

The time-evolution of the photon index for these bursts is shown in Fig.~\ref{fig:gammatime}.
The general behaviour of the bursts is similar to that seen in GRB~130925A, although the latter
is softer than the majority of even these bursts. The only burst with a softer
spectrum is GRB~060218, which was an atypical burst in which a strong thermal component
was detected, that evolved to lower temperatures \citep{Campana06}. It has also been suggested by
\cite{Sparre13} that GRB~100621A may have a thermal component; however, the presence of that component
is by no means certain, and appears to be limited to the early-time data, thus is unlikely to be the cause of the
late-time evolution we report here.

The afterglow light curve morphology of this collection of bursts is heterogeneous; with such a small sample
it is impossible to draw firm conclusions; however, the distribution of morphologies is similar to that reported
by \cite{Evans09} for the first 327 \swift-detected GRBs. This makes it unlikely that all of these GRBs
have late-time emission caused purely by dust with no contribution from an external shock, as we postulate for GRB~130925A, but dust scattering may
contribute to their emission. We therefore looked in the literature and GCN circulars for the 8 GRBs with spectral
softening (excluding GRB~060218) to see if the GRBs are reported either as being `dark' bursts 
(e.g.\ \citealt{Jakobsson04,vanderHorst09}) or significantly reddened bursts, both of which are likely indications of significant dust in the 
host galaxy. We found such evidence for 6 of the GRBs: GRB~080207 \citep{Kruhler12,Perley13}; GRB~081221 \citep{Melandri12};
GRB~090201 \citep{Melandri12}; GRB~090404 \citep{Perley13}; GRB~100621A \citep{Melandri12,Greiner13} and GRB~130907A \citep{Schmidl13}.
Additionally, \cite{Hunt14} reported significant dust in GRB~090417B. 
The remaining GRB (GRB~110709A) has only upper limits in the optical band, which may also indicate the presence of
dust. These results support a generalisation of our explanation for the spectral evolution of
GRB 130925A, namely that spectral softening of the X-ray afterglow of a GRB is the result of
dust scattering of the prompt emission. 

\begin{figure}
\begin{center}
\psfig{file=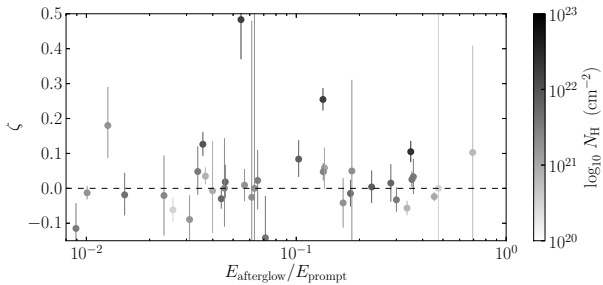,width=8.1cm}
\end{center}
\caption{The hardness ratio temporal evolution index ($\zeta$) as a function of the ratio of
prompt-to-afterglow energy release and intrinsic absorption. The ratio $E_{\rm afterglow}/E_{\rm prompt}$
refers to the integrated fluence of the afterglow and prompt models. If any objects were seen with a low
$E_{\rm afterglow}/E_{\rm prompt}$ ratio and either high intrinsic column and no spectral evolution; or spectral 
evolution but a low intrinsic column, this would contradict our model that spectral evolution is indicative 
of dust in the host galaxy. No such bursts are seen, supporting this model. Note that GRB 130925A is not
included in this plot.}
\label{fig:softtest}
\end{figure}

Note that this conclusion cannot necessarily
be inverted to argue that a highly extincted optical afterglow should correspond to a spectrally evolving
X-ray afterglow: this is only the case when the dust echo is of significant brightness relative
to the external shock, and the redshift is \sqiglt1.5 (at higher redshift the bulk of the dust echo fluence
lies below the XRT energy band).
We selected all GRBs within this redshift range, and plotted
the index of the HR evolution, $\zeta\pm\sigma_\zeta$, against the ratio, $E_{\rm afterglow}/E_{\rm prompt}$, 
coloured according to the intrinsic absorption column (according to the late time spectral fits in the XRT
Spectrum Repository\footnote{http://www.swift.ac.uk/xrt\_spectra}, \citealt{Evans09}). We searched 
for any examples with a high ($>10^{22}$ \cms) column and faint afterglow, but no evidence for spectral evolution; objects
which would argue against our interpretation. We found no such cases (Fig.~\ref{fig:softtest}).
We therefore suggest that the range of light curve morphologies seen in our sample of softening
afterglows indicates the differing relative strengths of the dust echo and external shock. 
GRB~130925A, with an exceptionally weak external shock (Section~\ref{sec:agboth}) is the most extreme example.

\section{Discussion}
\label{sec:disc}

GRB~130925A was a very long GRB, with high-energy emission ($E>15$ keV) detected until \til5 ks after the
initial trigger, and the prompt emission dominating the light curve until \til20 ks after the trigger.
Three other GRBs (101225A, 111209A and 121027A) also show such long-lived activity, prompting
some authors \citep{Gendre13,Levan14} to suggest that these belong to a new category of `ultra-long' GRBs. There is 
no formal definition of such objects, but the long duration of GRB~130925A clearly places it in this category.
These authors propose several possible causes of these ultra-long GRBs: most notably a tidal disruption event (TDE)
in which  a star is destroyed and partially accreted by a massive
black hole at the centre of a galaxy; and a GRB from the collapse of a blue supergiant (see also \citealt{Stratta13,Nakauchi13}), rather than the
Wolf-Rayet progenitor associated with `normal' long GRBs \cite{Woosley93}. However, the identification of these GRBs as 
a new class of object is not certain. Due to the low-Earth orbit of the \swift\ and \fermi\ satellites, it is 
difficult to accurately measure the duration of such long GRBs with these satellites. Indeed, for GRB~130925A 
we find that roughly 75\%\ of the fluence occurred during the second emission episode (\t0+2--3 ks; Section~\ref{sec:prompt}), which was completely missed
by \swift\ and \fermi. Similarly, for GRB~121027A a significant proportion of the emission
took place while \swift\ was not observing it (Starling et al., in prep), and for GRB~111209A the \kw\ light
curve\footnote{http://www.ioffe.rssi.ru/LEA/GRBs/GRB111209A/} shows that the emission continued for
about 3 ks after BAT finished observing. Thus, we cannot simply determine the distribution of 
GRB durations based on the \swift-BAT results.

\cite{Zhang14} attempted instead to define the duration of the burst as the maximum time over 
which emission from processes internal to the jet (i.e.\ prompt emission or X-ray flares) are seen. The distribution
of this duration has  broad long-duration tail, perhaps suggestive of a single population of
objects. \cite{Zhang14} suggested that this could be interpreted as indicating the duration of
the GRB central engine activity, which means that the GRB central engine is still active at the time
a flare is detected. Late-time X-ray flares \citep[e.g.\ ][]{Curran08} could, however, arise from internal shocks between two shells
of similar Lorentz factor,  in which case the time of collision could be much later than the time at which they were ejected
by the central engine; although \cite{Lazzati07} considered this scenario and suggested it was
more likely that the central engine was indeed still active at this time. Nonetheless, there is a
significant difference between these objects with late flares -- where the central engine apparently
turns off for a long period of time, and then emits a single, late-time flare -- and the ultra-long bursts 
where the central engine is active and highly energetic for a sustained period.

\begin{figure}
\begin{center}
\psfig{file=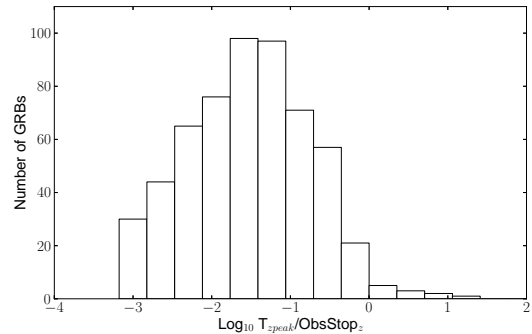,width=8.1cm}
\end{center}
\caption{The distribution of the time of GRB pulses ($T_{zpeak}$) relative to the time \swift\ slewed away from the burst
(ObsStop$_z$).
The GRBs in this plot are those from Figs.~\ref{fig:pulse1}--\ref{fig:pulse2}, with GRB 130925A excluded. The lack
of a sharp drop at $T_{zpeak}/$ObsStop$_z$=1 shows that the absence of late-time pulses in most long GRBs is not an observational
selection effect.}
\label{fig:slewtimes}
\end{figure}

GRB~130925A is the first of these ultra-long bursts to which the pulse modelling of \cite{Willingale10}
has been applied. Fig.~\ref{fig:pulse1} showed that while the durations of the individuals pulses lie within
with the distribution seen from the GRB population at large, the number of pulses and their peak times do not.
To determine whether this is the result of the selection biases referred to above (i.e.\ we cannot detect pulses when \swift\ is not
observing the burst), we plot in Fig.~\ref{fig:slewtimes} the distribution of the pulse times divided by the times at
which \swift's first observation  of the GRB ended. This shows values only for GRBs shown in Figs.~\ref{fig:pulse1}--\ref{fig:pulse2}
with GRB~130925A excluded. Whereas  Fig.~\ref{fig:pulse1} (centre panel) shows that the distribution of flare peak times
drops off sharply at around \t0+100~s, Fig.~\ref{fig:slewtimes} shows that there is no sharp drop corresponding to the
end of the \swift\ observation. It is highly improbable that GRBs systematically return to quiescence a few minutes after the trigger, and then flare
up again when \swift\ has slewed away (without triggering any other GRB satellite during
this later episode). We therefore suggest, given the lack of GRBs with peak times or numbers of pulses between those of GRB~130925A
and the bulk of the distribution, that the prompt emission of GRB~130925A and the other ultra-long bursts, are
not consistent with the tail of some continuous distribution of behaviours seen in ordinary long GRBs (as suggested by \citealt{Virgili13}).

\begin{figure*}
\begin{center}
\psfig{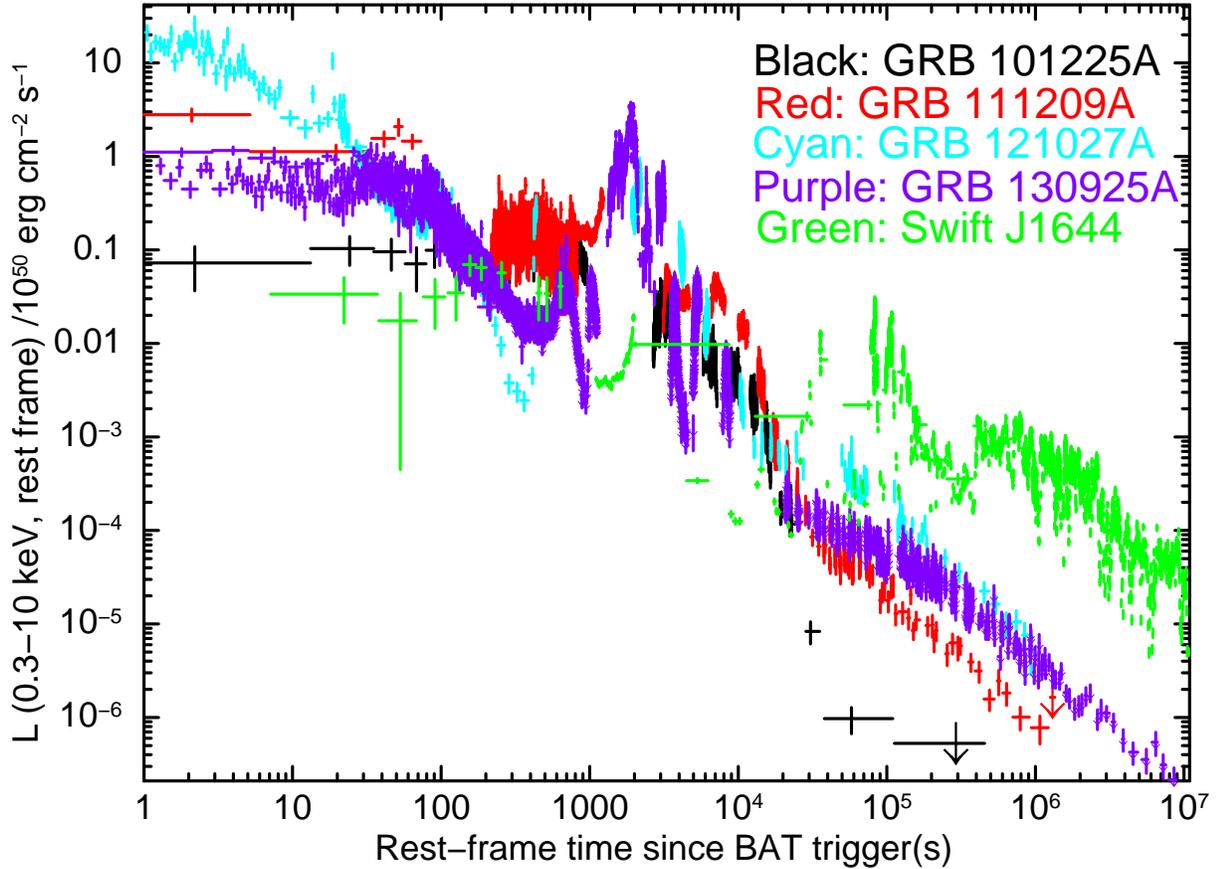}
\end{center}
\caption{The rest-frame X-ray light curves of the ultra-long GRBs identified by Levan \etal(2014), and GRB~130925A.
Swift J1644 is also shown for comparison. The energy band is 0.3--10 keV in the rest frame. For all but GRB~130925A
only the BAT event data and the XRT data are shown, $k$-corrected from the \swift\ Burst Analyser (Evans \etal2014)
and Light Curve Repository (Evans \etal2007, 2009) respectively. For GRB~130925A, the \kw\ data are also shown; these provide
the data around 1000--3000 s. The similarity between the 4 ultra-long bursts can be seen, as can the difference between
these and the TDE candidate Swift J1644.}
\label{fig:compare_lc}
\end{figure*}

In Fig.~\ref{fig:compare_lc} we show the \swift-XRT light curves of all four of the candidate ultra-long GRBs, converted 
to luminosity in the rest-frame 0.3--10 keV band. There are strong similarities between them, especially at around \t0+\til20 ks,
where the prompt emission appears to cease. We thus interpret GRB~130925A as belonging to the class of ultra-long GRBs and
that these are a distinct class of objects. We now consider the plausibility of
both the TDE and GRB scenarios for GRB~130925A.

\subsection{Is GRB~130925A a TDE?}
\label{sec:discTDE}

If GRB~130925A is a tidal disruption event (TDE), we would expect it to be located 
at the centre of the galaxy, where the supermassive black hole should lie. However, the \emph{HST} observations
show it to lie 0.12$^{\prime\prime}$ (\til600 pc) away from the galaxy nucleus \citep{Tanvir13}. Those same
observations show the galaxy to be somewhat distorted, suggestive of a recent merger; in such a case 
the galaxy could potentially host two such black holes which have not yet had time to merge and return to the
centre of mass \citep[e.g.\ ][]{Comerford13,Milosavljevic01}, thus the offset does not rule out the TDE scenario.

As \cite{Levan14} pointed out, a bigger problem faced by the TDE scenario is that of timescales: 
for disruption of a main-sequence star by a $10^6 \msol$ black hole \cite{Lodato11} predicted that
the X-ray emission would show a rise or plateau lasting \til100 days or more, whereas for GRB~130925A 
the light curve is steadily decaying by \til0.3 d after the trigger\footnote{It is worth noting that
Swift~J1644.3+573451, discussed shortly, is believed to be a TDE, but has a plateau of only \til 10 days; however, this is
still much longer than GRB~130925A}. \cite{Krolik11} considered
the case of a white dwarf being tidally disrupted by a lower-mass ($10^4 \msol$) black hole; as \cite{Levan14}
noted, their equation (5) represents the shortest timescale on which we may see variations. Equating this to
the \til2 ks gap between the burst episodes requires a 600 (2\tim{4}) \msol\ black hole for a 1 (1.4) \msol\
white dwarf. These values are not impossible, but clearly to explain the observed timescale in terms
of a TDE requires either a relatively low-mass black hole, or high mass white-dwarf.

The predicted peak brightness of TDEs is also a problem. \cite{Levan14} commented that the ultra-long GRBs are much more luminous (during their prompt
emission) than Swift~J1644.3+573451, which is believed to be a TDE detected by \swift\ \citep{Bloom11,Levan11,Burrows11}\footnote{As
\cite{Levan14} note, Swift~J1644 has a peak luminosity well above the predictions, but the ultra-long GRBs
have even more luminous peaks.},
and this is clear from Fig.~\ref{fig:compare_lc}. \cite{Lodato11} performed numerical
simulations of TDEs for a range of black hole masses, and report peak isotropic luminosities of $\til10^{44}$ erg s$^{-1}$
(see their fig.~7, for example). The average luminosity of GRB~130925A during the prompt phase is \til$E_{\rm iso}/2200$ (i.e. the prompt energy release
divided by the approximate `on time' of the burst) $\approx1.3\tim{50}$ erg s$^{-1}$, which is many orders of magnitude
above the predicted TDE peak. To reconcile these numbers by assuming that in GRB~130925A the radiation we see
is beamed requires a jet opening angle of \til0.07\deg. While this may not be impossible, it would mean that for every TDE we detect, 
about $10^7$ are beamed away from us. Given that \swift\ has detected 4 ultra-long GRBs in 9 years out to $z=1.773$, i.e.\ a volume
of 4.9\tim{5} Mpc$^3$, this implies a TDE rate of \til9 yr$^{-1}$ Mpc$^{-3}$, greatly in excess of the predicted rate
of 10$^{-5}$ yr$^{-1}$ Mpc$^{-3}$ \citep{Wang04}. Further, due to its shortness (for a TDE) such beaming reduces the 
overall fluence of the TDE to 2.2\tim{47} erg\footnote{Ignoring the later-time emission, which we showed in Section~\ref{sec:agboth}
to be negligible compared to the prompt emission}, which corresponds to the accretion of $10^{-6}$ \msol\ of material
(assuming 10\%\ radiative efficiency, \citealt{Lodato11}); whereas \cite{Ayal00} suggest that about 10\%\ of the
stellar mass will be accreted.

Another difficulty with the TDE scenario is the lack of fallback emission. 
Once the initial disruption event is over, some fraction of the stellar matter 
is accreted on to the black hole, producing a light curve which decays as $t^{-5/3}$ \citep[e.g.][]{Rees88,Phinney89,Evans89}.
However, the late-time emission in GRB~130925A is best modelled by dust scattering of the
early emission, not fallback emission. We therefore tried to determine limits on the possible emission from this fallback.
In Section~\ref{sec:agboth}  we determined the upper limit on the energy from a standard GRB afterglow to be 3.3\tim{50} erg.
Assuming 10\%\ radiative efficiency, this corresponds to the accretion of just \til2\tim{-3} \msol\ of
material, much lower than predicted by \cite{Ayal00}. Given that the standard afterglow
model we used to derive the limit on the afterglow emission decays more slowly than $t^{-5/3}$ (i.e.\ TDE decay),
the limit on emission from fallback accretion is even lower than 2\tim{-3} \msol.

We therefore consider it very unlikely that GRB~130925A can be explained in the TDE paradigm.

\subsection{Is GRB~130925A a GRB collapsar?}
\label{sec:discGRB}

There are two difficulties to interpreting GRB~130925A as a normal long GRB: its long-lived emission
at high energies ($E>15$ keV; Section~\ref{sec:pulsemodel})
and the low luminosity of the external shock emission (Section~\ref{sec:agboth}). The former is, by definition,
common to the ultra-long GRBs, and \cite{Gendre13}, \cite{Nakauchi13} and \cite{Levan14} have
suggested that it could be explained by the collapse of a blue supergiant, as opposed to the smaller Wolf-Rayet progenitor of normal long GRBs.
Considering the lack of external shock emission, Fig.~\ref{fig:compare_lc} shows that GRB~101225A also has little or no afterglow emission.
GRBs 121027A and 111209A have similar late-time X-ray light curves to
GRB~130925A, but there is no sign of spectral softening (the signature of dust scattering), implying that
in those bursts the X-ray emission arises from the standard external forward shock. However, the similarity of
their light curves with GRB~130925A tells us that the ratio of prompt-to-afterglow fluence
for those GRBs must be similar to GRB~130925\footnote{Here `afterglow' refers to the late-time X-ray emission,
rather than specifically external shock emission.}, i.e.\ all four
of the ultra long GRBs have afterglows which are under-luminous compared to their prompt emission, when compared with the population of normal long GRBs.

We now consider specifically the lack of external shock emission in GRB~130925A.
For a standard afterglow, the brightness of the external shock depends on the microphysical parameters of the
shock, which cannot be constrained by the XRT limit alone. Fortunately, the radio data from \cite{Bannister13}
at \t0+15 days, are close in time to the second-epoch HST data which give
$F_{1.6\mu m}\til0.6 \mu Jy$ (Tanvir et al., in preparation). From the \swift\ Burst Analyser \citep{Evans10}, the X-ray
flux density at 10 keV at this time was \til10$^{-4} \mu Jy$, with the contribution from the external shock being
at least a factor of three lower (Section~\ref{sec:agboth}). A rough SED constructed from these data
does not allow us to place stringent constraints on the afterglow properties, but is consistent
with a synchrotron model, where the electron distribution index $p=2.2$ (where $N(E)\propto E^{-p}$)
and $\nu_m<\nu_{\rm radio,HST}<\nu_c<\nu_x$ (where $\nu_m$ is the synchrotron peak frequency, $\nu_c$ is the
cooling frequency, and $\nu_{\rm radio,HST,x}$ are the frequencies of the radio, HST and XRT emission respectively).
Using the equations of \cite{Granot02} this loose constraint on $\nu_c$ gives
$10^{-4}$ \sqiglt\ $n$\ \sqiglt$1.5$ cm$^{-3}$, but also predicts an X-ray flux significantly
higher than measured. 
In order to bring the predicted flux into agreement with the
observations we have to reduce the kinetic energy of the outflow to \til 5\tim{51} erg.
Alternatively, we can in principle
suppress the flux if the magnetic parameter of the shock, $\epsilon_B$, is very low \citep[e.g.][]{Uhm07}; however,
in order to keep $\nu_c$ between the optical and X-ray bands while reducing $\epsilon_B$
requires the circumburst density to increase, and only unphysical values of $\epsilon_B$ and $n$ 
can reproduced the observed fluxes.

An alternative explanation is that the optical and radio emission comes not from the 
external forward shock, but from a reverse shock \citep{Hascoet11b,Uhm07,Genet07}. To fit the rough
SED we produced above, we again require $\nu_m<\nu_{\rm radio,HST}<\nu_c<\nu_x$, but in this case
the normalisation of the SED depends on the distribution of densities and Lorentz factors behind the shock
(see the papers just cited for details); the modelling of which is beyond the scope of this paper.
For the reverse shock model to work, it is still necessary to suppress the emission from the forward shock. 
The authors above do this by requiring $\epsilon_B$ to be low (\til$10^{-7}$) in the external shock.
Unlike in the situation described above for the forward shock, this is attainable because we have no observational constraints on the shape of the spectrum
from the (suppressed) forward shock.

Thus, if GRB~130925A is a GRB, we need to explain either why it should radiate a greater
proportion than normal of its energy during the prompt phase, 
or have an unusually low magnetic energy in the external shock. As noted above, the low luminosity of the
afterglow compared to the prompt emission appears to be common to all four ultra-long GRBs.
This raises the possibility that some mechanism related to the burst duration also increases the 
fraction of energy radiated during the prompt phase;
that is the fraction of the energy in the outflow which is converted to radiation.
It is tempting to interpret the bottom panel of Fig.~\ref{fig:pulse2} -- which shows that the pulses
in GRB~130925A tend to be longer lived for their luminosity than the general population of pulses -- as supporting this
idea. However, this does not tell us anything about the efficiency with which the energy contained in
the interacting matter is radiated.
The fraction of the intial energy radiated as prompt emission depends not only on the mechanism by which interactions in the outflow
dissipate energy, but also on how much of the outflow is involved in such interactions. In the standard internal shock model,
interactions occur when two shells of are material emitted at times $t_2>t_1$ with Lorentz factors $\Gamma_2>\Gamma_1$;
\emph{provided that the second shell catches up with the first one before the former is decelerated by the ISM at the external shock.}
Therefore prompt pulses can only be produced by shells which collide within $\til R_d/c$ s after being ejected, where $R_d$ 
is the deceleration radius of the shock; this increases with time as the shock propagates, but much more slowly
than the pre-shock outflow, thus at early times we can treat $R_d$ as \til constant.

This naturally predicts some limit to the apparent duration of the GRB, as pulses that would take longer than $\til R_d/c$ to interact
never do so and are thus not seen; instead the energy contained in those pulses is given to the external
shock. Thus, if pairs of shells with collision radii $>R_d$ are habitually emitted, we would expect
a cutoff in the distribution of GRB durations corresponding to $\til R_d/c$ and evidence for 
energy injection into the external shock after this time. Both of these exist: the former is seen in 
the central panel of Fig.~\ref{fig:pulse1} (cf Section~\ref{sec:disc}); the latter is the `plateau'
phase seen in X-ray GRB afterglows \citep[e.g.][]{Nousek06,Zhang06,Liang07}. Variations 
in the duration of central engine activity, the distribution of Lorentz factors it emits, the 
energy emitted and the density of the circumburst medium will all affect these signatures; broadening the
cutoff in duration and giving a range of plateau luminosities (including no plateau at all, if the engine emits
no pair of shells that collide after the deceleration radius). These are significant unknowns; we cannot 
quantitatively compare this prediction with the data, but they are at least qualitatively consistent.

In terms of the ultra-long GRBs: the presence of prompt pulses extending to such late times\footnote{These are
distinct from the late-time XRT flares occurring days after the trigger, on which timescales we cannot treat
$R_d$ as constant} compared to most bursts (Fig.~\ref{fig:pulse1}, middle panel) implies either that the central engine
continues to emit pairs of shells with $\Gamma_2\gg\Gamma_1$ (i.e. shells which interact close to the central engine)
for much longer than normal, or that the deceleration radius in those bursts is larger than normal,
allowing more of the emitted shells to interact before encountering the external shock; this is supported
by the top panel of Fig.~5, which shows that GRB 130925A had many more pulses, i.e. internal collisions,
than the normal GRBs. The decay timescale of a pulse is a function of the distance from the central engine
 at which the shells collide,
because the decay is caused by high latitude emission and the surface of the jet is larger (hence the high latitude emission longer)
at greater radii from the central engine.  The top panel of Fig.~\ref{fig:pulse2} shows that, in GRB~130925A, the late pulses are longer
in duration than earlier pulses in the GRB population at large, indicating that these late pulses are occurring at larger
radii than normal. This indicates that the deceleration radius in the ultra long GRBs is
larger than in normal GRBs. The increased number of pulses means that more of the initial 
energy is radiated away as prompt emission, simply because there are more processes to dissipate energy than
in a normal long GRB.

Our pulse modelling shows that the total energy output of GRB~130925A (and, by analogy to Fig.~\ref{fig:compare_lc},
the ultra long bursts generally) is not higher than in the 
general population of long bursts, so an increased $R_d$ implies a lower circumburst density -- as allowed
by our rough SED modelling above. The combination
of this lower density and the fact that more of the outflow is involved in dissipative internal shocks
implies that ultra-long GRBs should have 
$E_{\rm afterglow}/E_{\rm prompt}$ values lower than the normal long GRBs as we have found for GRB~130925A.
We have argued qualitatively that this is the case, based on Fig.~\ref{fig:compare_lc}, but
we can test this prediction in more detail. To do this, we fitted the \swift\ and \kw\ data of
GRB~121027A in a manner analogous to that in Section~\ref{sec:pulsemodel}, and found the prompt
fluence to be 1.6\tim{54} erg, while the afterglow fluence was 1.5\tim{52} erg. Comparison
with Fig.~\ref{fig:promptag} shows that, as predicted for the ultra-long GRBs, $E_{\rm afterglow}/E_{\rm prompt}$
for GRB 121027A (the cyan point) is notably lower than the population of bursts as a whole, supporting our model.
We also note that, under the ICMART model \citep{Zhang10} for prompt emission it is possible to get significant
variations in the efficiency with which internal shock interactions convert the kinetic
energy to radiation (e.g. \citealt{Zhang14}) which may also contribute; however, in this model
the interactions still have to occur inside the deceleration radius. An additional implication of our model of an increased
deceleration radius in the ultra-long GRBs
is that the ultra-long GRBs are unlikely to show a strong plateau phase in the afterglow. This is because the ejected shells of material,
which refresh the external shock to cause the plateau in a normal GRB, have dissipated some of their 
energy by internal shocks before reaching the external shock. This lack of plateau is consistent with the 
observations (Fig.~\ref{fig:compare_lc}) \cite{Stratta13} find evidence for a plateau in their \emph{XMM-Newton\/}
observations on GRB~111209A; however, as they note, it is one of the weakest plateaux observed, consistent with our model.

If our idea is correct, we do not require a different progenitor from ordinary long GRBs in order to explain
the burst duration, as previous works \citep{Gendre13,Nakauchi13,Levan14} have suggested. However,
we do require a low-density medium around a star massive enough to form a GRB,  which the 
low-metallicity blue supergiant model those authors propose would naturally explain.

\section{Conclusions}

GRB~130925A was an extremely long GRB at $z=0.348$, with an observer-frame duration of around 20 ks, and three main episodes
of emission at $E>15$ keV. Apart from its length, the properties of the prompt emission appear consistent with those
of other bursts. However, the extreme duration of this burst is inconsistent with the general population,
and we have ruled out observational bias as the cause of this incompatibility.

The late-time X-ray data show a strong spectral evolution, which can be well modelled as dust scattering
of the prompt emission. A systematic study of other GRBs shows evidence for such emission
in at least 8 other objects. GRB~130925A is the most extreme example, because in addition
to the dust echo, it shows no evidence for a contribution from a standard afterglow; we place 
a limit of $E_{\rm afterglow}<3.3\tim{50}$ erg, a factor of 1,000 lower than the energy released 
in the prompt phase. This faint (or missing) external shock is essential to the detection of a dust echo, 
because an external shock of normal brightness will otherwise outshine the echo.

We have considered two possible scenarios to explain this object: a tidal disruption event, or a
GRB.  The former is difficult to reconcile with the observed timescales, although the disruption of a 
white dwarf may be permissible if the masses are finely tuned. The energetics, and the lack of 
emission detected from fallback accretion, appear to rule out a TDE origin for GRB 130925A.

The lack of a standard, external-shock afterglow presents a challenge for the GRB interpretation, and even in a low density
environment ($n\til10^{-3}$ cm$^{-3}$) the ratio of the prompt fluence to the limit on the afterglow fluence can 
only be explained if the prompt emission process converts more of its energy to radiation than is typical
for GRBs. However, we argue that this is to be expected in  a low density circumburst medium, in which the external shock
forms at a greater distance from the GRB than normal, allowing more internal shocks to occur and dissipate energy
which, in a typical GRB, would instead be injected into the external shock. The ultra-long GRBs detected so far 
show a lower ratio of afterglow to prompt fluence than the population of normal long GRBs, supporting the idea that they occur
in a low-density environment.

\section*{Acknowledgements}

This work made use of data supplied by the UK Swift Science Data Centre at the University
of Leicester. PAE, JPO, KW and APB acknowledge UK Space Agency support.
The \emph{Konus-WIND}\/ experiment is partially supported by a Russian
Space Agency contract, RFBR grants 12-02-00032a and 13-02-12017 ofi-m.
DNB and JAK acknowledge support from NASA contract NAS5-00136.
This work includes observations made with the Gran Telescopio Canarias (GTC),
installed in the Spanish Observatorio del Roque de los Muchachos of the
Instituto de Astrof\'{\i}sica de Canarias, in the island of La Palma. This work was partially supported by the Spanish Ministry project
AYA2012-29727-C03-01.

\bibliographystyle{mn2e}
\bibliography{phil}

\label{lastpage}
\end{document}